
\magnification=1200
\hsize=6.75truein
\hoffset=-.125truein
\vsize=9.2truein
\voffset=-.3truein
\tolerance 600
\font\ninerm = cmr9
\def\svec#1{\skew{-2}\vec#1}
\overfullrule=0pt

\let\oldeqno = \eqno
\def\eqno#1$${\oldeqno{\hbox{#1}}$$}

\catcode`@=11
\def\eqalignno#1{\displ@y \tabskip\centering
  \halign to\displaywidth{\hfil$\@lign\displaystyle{##}$\tabskip\z@skip
    &$\@lign\displaystyle{{}##}$\hfil\tabskip\centering
    &\llap{$\@lign{\hbox{##}}$}\tabskip\z@skip\crcr
    #1\crcr}}
\catcode`@=12 

\def\footnoterule{\kern-3pt \hrule width \hsize \kern2.6pt}
\pageno=0
\footline={\ifnum\pageno>0\hfil--\folio--\hfil\else\hfil\fi}

\baselineskip 14pt plus 1pt minus 1pt
\centerline{\bf POLARIZED DEFORMED NUCLEI STUDIED VIA COINCIDENCE }
\centerline{{\bf POLARIZED ELECTRON SCATTERING:
The case of $^{21}{\overrightarrow{\hbox{Ne}}}$}\footnote{*}
{\ninerm This work is supported in part by funds
provided by the U. S. Department of Energy (D.O.E.) under contract
\#DE-AC02-76ER03069, and by DGICYT (Spain) under contract No.
PB92/0021--002--01.}}
\vskip 16pt
\centerline{J. A. Caballero\footnote{$^1$}
{\ninerm Present address: Instituto de Estructura de la Materia, CSIC,
Serrano 123, Madrid 28006, Spain}, T. W. Donnelly and
G. I. Poulis\footnote{$^2$}
{\ninerm Present address: TRIUMF, Vancouver, British Columbia,
Canada V6T 2A3}}
\vskip 8pt
\centerline{\it Center for Theoretical Physics}
\centerline{\it Laboratory for Nuclear Science}
\centerline{\it and Department of Physics}
\centerline{\it Massachusetts Institute of Technology}
\centerline{\it Cambridge, Massachusetts\ \ 02139\ \ \ U.S.A.}
\vskip 16pt
\centerline{E. Garrido and E. Moya de Guerra}
\vskip 8pt
\centerline{\it Instituto de Estructura de la Materia}
\centerline{\it Consejo Superior de Investigaciones Cient\'{\i}ficas}
\centerline{\it Serrano 123, Madrid\ \ 28006\ \ \ SPAIN}
\vfill
\centerline{\bf ABSTRACT}
\medskip
{\narrower
Coincidence reactions of the type $\svec{A}(\svec{e},e'N)B$ involving the
scattering of polarized electrons from deformed polarized targets are
discussed within the context of the plane--wave impulse approximation. A
general expression for the polarized spectral function for transitions leaving
the residual nucleus in discrete states is presented. General properties and
angular symmetries exhibited by the polarization observables are discussed in
detail. Results for unpolarized cross sections as well as for polarization
ratios (asymmetries) are obtained for typical quasi--free kinematics. The
dependences of the polarization observables on the bound neutron momentum,
target polarization orientation, nuclear deformation and value of the momentum
transfer $q$ are discussed in detail for various different kinematical
situations. \smallskip}
\vfill
\centerline{Submitted to: {\it Nuclear Physics A}}
\vfill
\line{CTP\#2098 \hfill hep-ph/9308221 \hfill July 1993}
\eject

\baselineskip=12pt plus1pt minus1pt

\noindent{\bf 1.\quad INTRODUCTION}
\medskip\nobreak
This work is based on a study of polarization observables in coincidence
electron scattering reactions within the context of the Plane--Wave Impulse
Approximation (PWIA) or factorized Distorted--Wave Impulse
Approximation (DWIA), where the cross section can be
factorized into two basic terms, the electron--nucleon cross section and the
spectral function.$^{[1-4]}$  We shall consider only the PWIA in detail in the
present work.
The former factor deals directly with the interaction between the
incident electrons and the bound nucleons inside the nucleus, while the latter
gives us the probability that a nucleon is to be found in the nucleus with
given energy and momentum. Although the PWIA is evidently an
oversimplification in the description of the reaction mechanism, it
gives us a very clear picture of the physics contained in the problem and
has proven to be quite useful in studying the single--nucleon content of the
nucleus under appropiate kinematical conditions (specifically, high enough
values of $q$ where the scattering process is expected to be
``quasi--free'' and hence only mildly influenced by final--state interactions
and various exchange effects which are ususally neglected or, at best, only
treated approximately$^{[1,2]}$).

The study of the electron--nucleon cross section for polarized incident
electron and polarized target has already been presented in detail in
Ref.~[3].  A very important advantage of the PWIA is the ability to
treat some of the relativistic aspects of the reaction in a complete
way. At high values of the momentum transfer and quasi--free kinematics,
the ejected nucleon becomes relativistic and hence any fully
non--relativistic treatment of the reaction must be viewed with
caution.  In this paper we place our focus on the second of the factors
above, namely, on the nuclear spectral function. Our approach for this
quantity is non--relativistic; however, typically the most important
contributions to the cross section come from struck nucleons within the
Fermi sea ($p <p_F\approx 200-250$ MeV/c) and consequently
non--relativistic approximations for the initial--state structure (and
therefore for the spectral function) may be expected to be approximately
valid.

As previously stated, the analysis within the context of the PWIA/DWIA of
coincidence polarized electron scattering from polarized nuclei
leads to factorized expressions for the differential cross sections.
These involve the {\it polarized} single--nucleon
cross section$^{[3]}$ multiplied by a {\it polarized} spectral function.
The latter provides the probability of finding a nucleon in the nucleus having
given energy and momentum and, importantly, given spin projection.
Specifically,
we consider in detail the case of particular transitions (to ground and
low--lying excited states) involving the polarized nucleus
$^{21}{\overrightarrow{\hbox{Ne}}}$. This
target constitutes an interesting case, being a deformed nuclear system with
a well--established rotational spectrum$^{[5]}$ and yet is light enough that
electron distortion effects can be neglected, at least at this early stage
of the theoretical analysis. Of considerable importance for practical
experiments is the fact that, as for $^3{\overrightarrow{\hbox{He}}}$,
the case of
polarized neon involves a noble gas and hence depolarization effects from
collisions with target--container walls are minimized, permitting
high--density polarized targets to be constructed.

Using the nuclear rotational model of Bohr and Mottelson$^{[6]}$ to
describe polarized neon, the present studies provide a starting point
for investigating spin degrees of freedom in coincidence electron
scattering from such nuclei; our present interest is focused on
providing a guide to where the greatest sensitivities to particular
aspects of the polarized nuclear spectral functions are to be found.  We
present results for asymmetries and/or polarization ratios as well as
for totally unpolarized cross sections. In so--doing we stress the
potential importance of studying ratios of observables which turn out to
be less sensitive to the nature of the underlying dynamical assumptions
made (at least they will be shown to have different sensitivities).  We
have examined the behaviour of the asymmetries as functions of the
struck--nucleon momentum $p$ as well as versus the angles defining the
direction of the target polarization vector for different
kinematics. The effects introduced into the different responses by the
deformation of the target and its connection with the polarizations, as
well as the behaviour of the various responses in regions where
relativistic effects in the currents may be expected to play a
significant role ($q=1$ GeV), have been also considered and the results
are discussed in detail in the present work.

The study of deformed nuclear systems through electron scattering
reactions started some years ago.$^{[7,8]}$ In the particular case of
inclusive (single--arm) processes, the predictions of various models for
the different components of the electromagnetic current of the
ground--state band in even--even and odd--$A$ deformed nuclei in the
rare--earth region have been already presented in previous
papers.$^{[9,10]}$ These ideas have also been applied to lighter nuclear
systems where one can neglect the Coulomb distortion effects which
otherwise complicate the analysis;$^{[11]}$ in particular, the case of
$^{21}$Ne has been treated in detail in Ref.~[12]. For coincidence
reactions of the type $A(e,e'N)B$, a great deal of effort has been
expended in recent years to measure momentum distributions in nuclei. An
interesting question that arises is whether momentum distributions in
deformed nuclei may look different from those in the spherical case,
{\it i.e.,\/} whether nuclear deformation may lead to observable effects
in coincidence quasielastic electron scattering reactions. This subject
was first studied in the totally unpolarized situation (incident
electron and target not polarized) some years ago.$^{[8,13-15]}$ In this
work our interest has been to generalize such studies when polarization
degrees of freedom are taken into account and to provide a new probe of
the deformed, spin--dependent spectral function of rotational nuclei. In
the future we intend to explore some natural extensions of the present
work --- in particular to include treatments beyond the PWIA and to
describe polarized coincidence reactions at large inelasticity.

This paper is organized as follows: in Sect.~2 we present a brief
summary of the general formalism needed in treating the process
$\vec{A}(\vec{e},e'N)B$ within the context of PWIA, including a summary
of the basic ingredients entering into the single--nucleon response
functions drawn from our previous work.$^{[3]}$ The calculation and
discussion of the general properties of the polarized spectral function
for deformed nuclear systems are discussed in Sect.~3 (with some
developments given in Appendix A) and in Sect.~4 we present and analyze
the general expressions obtained for the asymmetries. Results are
presented in Sect.~5: specifically, details concerning the description
of the target $^{21}$Ne are presented in Sect.~5.1, while in Sect.~5.2
we discuss the kinematics used in the calculations, and then in
Sects.~5.3--5.6 results for the cross sections, asymmetries and response
functions are presented.  Finally in Sect.~6 we give a summary of what
has been learned from these investigations and present our conclusions.

\goodbreak\bigskip
\noindent{\bf 2.\quad COINCIDENCE ELECTRON SCATTERING FORMALISM}
\medskip\nobreak
\noindent{\bf 2.1\quad General Exclusive Electron Scattering}
\medskip\nobreak

Following Refs.~[3,16], in this section we briefly summarize the essential
parts of the general formalism involved in describing reactions
of the type $\vec A(\vec e,e'N)B$, where incident electron and
target nucleus $(A)$ are polarized. We restrict our attention
to the extreme relativistic limit
(ERL) for the electrons and to the case where the residual nucleus $B$ is
left in a bound state.
The process in the Born approximation
is represented using the Feynman diagrams in Fig.~1. The different
kinematic variables in the laboratory frame are the following:
$K^{\mu} \equiv (\epsilon, {\bf k})$
and $K'^{\mu}\equiv (\epsilon ',{\bf k'})$
are the four--momenta of the incident
and scattered electrons, respectively, while the
hadronic variables $P_A^{\mu}\equiv (M_A,{\bf 0})$,
$P^{\mu}_B \equiv (E_B , {\bf p_B})$, $P^{\mu}_N \equiv (E_N ,{\bf p_N})$ are
the four--momenta of the target, residual nucleus and emitted nucleon in the
laboratory frame, respectively.
We have the relationships $E_B=\sqrt{p_B^2
+ M_B^2}$ and $E_N=\sqrt{p_N^2 + M_N^2}$, where $M_N$ is the nucleon
mass, $M_B$ is the rest--mass of the residual nucleus (and includes
any internal excitation energy in that system), $p_B\equiv |{\bf p_B}|$
and $p_N\equiv |{\bf p_N}|$.
The four--momentum transferred by the virtual photon is given
by $Q^{\mu}\equiv (\omega,{\bf q}) = (K-K')^{\mu} = (P_N+P_B-P_A)^{\mu}$.
Here the energy transfer is $\omega=\epsilon-\epsilon'$ and the
three--momentum transfer is ${\bf q}={\bf k}-{\bf k'}$, with the magnitude of
the latter being denoted $q\equiv |{\bf q}|$.

Using the property of
conservation of the nuclear electromagnetic current and
after integrating over the nucleon energy $E_N$
one can write the general expression for
the differential cross section for exclusive electron scattering:$^{[3,4,16]}$
$$\eqalign{
{d\sigma^h\over d\epsilon ' d\Omega _e d\Omega _N} &=
{p_N M_N M_B \over (2\pi)^3 M_A}\sigma_{\rm Mott} f_{\rm rec}^{-1}
\left\{R+hR'\right\} \cr
&\equiv\Sigma + h \Delta\ \ ,\cr}\eqno(1)$$
containing the helicity--sum (electron unpolarized)
and helicity--difference (electron polarized) cross sections, $\Sigma$
and $\Delta$, respectively. The other quantities in Eq.~(1) are defined
in Ref.~[3,16].
Note in the above expression that integration over the emitted
nucleon energy $E_N$ actually means that the value of the momentum $p_N$ is
completely specified by solving the energy balance equation
$$\sqrt{p_N^2+M_N^2}+\sqrt{p_N^2-2p_Nq\cos\theta_N+q^2+M_B^2}=
M_A+\omega\ \ , \eqno(2)$$
with $\theta_N$ the relative angle between the momenta ${\bf p}_N$ and
${\bf q}$.
The functions $R$ and $R'$ represent the hadronic responses
involving unpolarized or polarized electrons, respectively.
Both classes of responses may in general have contributions due to the
orientation of the target. They can be decomposed, as usual, into six
general classes of response labelled $L$, $T$, $TL$ and $TT$ for unpolarized
($T'$ and $TL'$ for polarized) electron scattering. Each response is
multiplied by its corresponding lepton kinematical factor, $v_L$, $v_T$,
{\it etc.\/}, where expressions for the six response functions $R^K$ and
kinematical factors $v_K$
are given in Ref.~[16]. We consider the case where only the target and/or
the initial
electrons are polarized, while final polarization is not observed.
It can be shown
that the dependence on the azimuthal angles $\phi_N$ and $\phi^*$
corresponding to the emitted nucleon and target polarization directions,
is the following:$^{[17]}$
$$\eqalign{\Sigma\sim & v_LW^L(\Delta\phi)
+ v_TW^T
(\Delta\phi)\cr
&+v_{TL}\left(\cos\phi_N W^{TL}(\Delta\phi)
+\sin\phi_N \tilde{W}^{TL}(\Delta\phi)\right)\cr
&+v_{TT}\left(\cos 2\phi_N W^{TT}(\Delta\phi) +
\sin 2\phi_N \tilde{W}^{TT}(\Delta\phi)\right)\cr}\eqno(3a)$$
and
$$\eqalign{\Delta \sim & v_{T'}
\tilde{W}^{T'}(\Delta\phi)\cr
& + v_{TL'}\left(\sin\phi_N W^{TL'}(\Delta\phi)
+ \cos\phi_N \tilde{W}^{TL'}(\Delta\phi)\right)\ \ ,\cr}
\eqno(3b)$$
where each response depends on $q$, $\omega$, $p_N$ and $\theta_N$ as
well as on the target polarization angles $\theta^*$  and $\Delta\phi$.
It should be noted that these results have been expressed in terms of
$\Delta\phi\equiv\phi^*-\phi_N$ (see Fig.~2).

In the particular
case of the PWIA to be discussed below, the terms $\tilde{W}^{TL}(\Delta\phi)$
and $\tilde{W}^{TT}(\Delta\phi)$ do not appear. Furthermore, for
situations where the target nucleus is unpolarized,
all responses with tildes vanish;
in PWIA the response $W^{TL'}(\Delta\phi)$ also vanishes.
All of the comments made in the case of PWIA can be also applied to the
factorized DWIA and the fact that certain (time--reversal odd, see Refs.~[17])
responses are absent in the PWIA or factorized DWIA reflects the nature of
these approximations. To the extent that experimental studies yield nonzero
results for these responses, it will be possible to evaluate the quality of
these (factorized) approximations. In Sect.~4 the dependence on
$\Delta\phi$ of the different responses within the PWIA will be made explicit.

\goodbreak\bigskip
\noindent{\bf 2.2\quad The Plane--Wave Impulse Approximation}
\medskip\nobreak
{}From now on we will focus on coincidence electron scattering within the
context of the PWIA. Here, as is well known,$^{[1-4]}$ in addition
to restricting the currents to one--body operators (Impulse
Approximation), one makes several
more stringent assumptions. Firstly,
one takes the emitted nucleon to be a plane wave, {\it i.e.,\/}
the nucleon is ejected from the nucleus without any further interaction
with the residual nuclear system. Secondly, one assumes that the nucleon
detected in the coincidence reaction is the one to which the virtual photon
is attached (see Fig.~3) and thus neglects various classes of exchange effects.
Within the PWIA, the general expression for the six--fold differential
cross section can be written as the following:$^{[3]}$
$$\eqalignno{
{d\sigma\over d\Omega_e d\epsilon' d\Omega_N dE_N} &=
{p_N M_N M_B\over E_B}\
\sum_{m m'} \sigma_{m m'}^{eN}
S_{m m'}({\bf p},E, \Omega^*)  &(4a)\cr
&= p_N E_N \sum_{m m'} {\tilde\sigma}_{m m'}^{eN}
{\tilde S}_{m m'}({\bf p},E, \Omega^*)
\ \ , &(4b)\cr}$$
where $\sigma_{m m'}^{eN}$ is the
``off--shell polarized electron--nucleon cross section''
and $S_{m m'}({\bf p},E,\Omega^*)$ the polarized spectral function
whose diagonal components ($m=m'$)
give the probability to find a nucleon in the target with
momentum ${\bf p}$, energy $E$ and spin projection $m$.
For convenience and to connect to previous work$^{[4]}$ in Eq.~(4b)
we have introduced a secondary single--nucleon cross section and a
secondary spectral function via the following:
$$\eqalignno{{\tilde\sigma}_{m m'}^{eN} &\equiv \Bigl(
{{M_N^2}\over{{\bar E}E_N}}\Bigr)\sigma_{m m'}^{eN} &(5a)\cr
{\tilde S}_{m m'}({\bf p},E, \Omega^*)  &\equiv \Bigl( {{M_B}\over{E_B}} \Bigr)
\Bigl( {{\bar E}\over{M_N}} \Bigr) S_{m m'}({\bf p},E, \Omega^*)  \approx
S_{m m'}({\bf p},E, \Omega^*)
\ \ , &(5b)\cr}$$
where ${\bar E}\equiv \sqrt{p^2+M_N^2}$.  The energy of the struck nucleon
is given by $E=M_A-E_B$ and in general ${\bar E}\neq E$, namely, the
kinematics are off--shell for this particle. Another energy commonly
used in discussions of coincidence electron scattering is the missing energy,
defined as $E_m \equiv M_N+M_B-M_A
=\omega-T_N-T_B$, where $T_N$ and $T_B$ are the kinetic energies
of the ejected nucleon and the (excited) residual nucleus, respectively.
Clearly one may use $E$ or $E_m = M_N+M_B-E_B-E$ as an argument in the
spectral function --- in this work we shall use $E_m$ rather than $E$.
Alternatively, it has proven useful$^{[3]}$ to use still
another energy ${\cal E}\equiv E_B-E_B^0$, where $E_B^0$ is the
energy of the daughter nucleus {\it in its ground state}. This has the merit
of involving a simple bound, {\it viz.\/} ${\cal E}\geq 0$, by construction.
Naturally one can write expressions for $E=E({\cal E},p)$ and
$E_m=E_m({\cal E},p)$ and thus can write $S_{mm'}$ as a function of
${\cal E}$ and $p$.

Of course, in coincidence electron scattering the
kinematics demand a specific relationship between the energies.  Specifically,
in specifying the electron scattering kinematics $q$ and $\omega$ are
fixed; detecting a nucleon amounts to fixing $p_N$ (or $E_N$) and
$\theta_N$ and hence, through Eq.~(2), $M_B$ is specified.  Finally,
since $p=\sqrt{p_N^2-2p_N q\cos\theta_N+q^2+M_B^2}$, the struck nucleon
momentum $p$ is also specified and therefore ${\cal E}$, $E$ and $E_m$
are as well. Introducing the binding energies of the target and daughter
nucleus, $\epsilon_A^o\equiv A M_N-M_A$ and $\epsilon_B\equiv (A-1)M_N
-M_B$, respectively,
the expressions for $\sigma_{m m'}^{eN}$ and
$S_{m m'}({\bf p},E_m,\Omega^*)$ are given by$^{[3]}$
$$\sigma_{m m'}^{eN}={2\alpha^2\over Q^4} \Bigl({\epsilon '\over
\epsilon}\Bigr)\eta_{\mu \nu} {\cal W}_{m m'}^{\mu \nu}({\bf p};{\bf q})
\ \ .\eqno(6)$$
and
$$S_{m m'}({\bf p},E_m,\Omega^*)=
\sum_{A}p(A) \sum_{B}
\langle B|a_{{\bf p}m'}|A\rangle ^{*}
\langle B|a_{{\bf p}m}|A\rangle\delta (E_m
+\epsilon_B-\epsilon_A^o)\ \ ,\eqno(7)$$
where $\eta_{\mu \nu}$ is the leptonic tensor (in the ERL with only the
incident electron polarized) and
${\cal W}_{m m'}^{\mu \nu}({\bf p};{\bf q})$ is the single--nucleon
tensor that depends on the $\gamma NN$ vertex (see discussion in
Ref.~[3]). Since we assume that no final polarizations are measured,
in the spectral function the
sum over $B$ involves all possible nuclear states including a sum
over magnetic substates. On the other hand,
the initial state is assumed to be polarized and this is represented by the
sum over $A$ with the weighting factor $p(A)$, {\it viz}., the probability
that specific projections of the ground--state angular momentum occur; here
$\Omega^*=(\theta^*,\phi^*)$ represents the angular variables defining the
target polarization direction (see Fig.~2).

The cross section after integrating over the energy $E_N$ can be expressed as
$$\eqalign{
{d\sigma\over d\Omega_e d\epsilon' d\Omega_N} &=
{p_N M_N M_B\over M_A}\ f_{rec}^{-1}
\sum_{m m'}  \sigma_{m m'}^{eN}
n_{m m'}({\bf p}, \Omega^*)\ \ ,\cr}\eqno(8)$$
with
$$n_{m m'}({\bf p},\Omega^*)=
\sum_{A}p(A) \sum_{B}
\langle B|a_{{\bf p}m'}|A\rangle ^{*}
\langle B|a_{{\bf p}m}|A\rangle\ \ ,\eqno(9)$$
the spin--dependent density matrix in momentum space.
The single--nucleon tensor and the spectral function
(and hence, $n_{m m'}({\bf p},\Omega^*)$)
can easily be shown to satisfy the following general symmetries:
$$\eqalignno{
{\cal W}^{\mu\nu}_{m m'}({\bf p};{\bf q})
&={\cal W}^{*\nu\mu}_{m' m}({\bf p};{\bf q})  &(10a)\cr
S_{m m'}({\bf p},\bar{E},\Omega^*) &=
S^*_{m' m}({\bf p},\bar{E},\Omega^*)\ \ .&(10b)\cr}$$
In Eqs.~(8,9) one should note that
when polarization degrees of freedom are taken into
account, in contrast with the unpolarized case,
the spectral function and electron--nucleon cross section
in general will contain both diagonal and off--diagonal spin components.
The various elements that make up the PWIA (and also the DWIA)
descriptions of
$\vec A(\vec e,e'N)B$ reactions
can all be considered to be $2\times 2$ hermitian matrices in spin--space,
where the four components
correspond to all the possible values of the projections of the spin of
the bound nucleon, $m, m'=\pm1/2$ along a specific direction.

Imposing current conservation for the single--nucleon current
the ``off--shell polarized electron--nucleon cross section'' can be decomposed
in the following way:
$$\sigma^{eN}_{m m'}= \sigma_{\rm Mott}\left\{\sum_{K}v_K
{\cal R}^K_{m m'}+h\sum_{K'}v_{K'}
{\cal R}^{K'}_{m m'}\right\}\ \ ,\eqno(11)$$
where the electron kinematical factors ($v_K, v_{K'}$)
and single--nucleon response functions ${\cal R}^{K/K'}_{m m'}$
are labelled as usual by $K=L, T, TL, TT$ and $K'= T', TL'$.
The response functions are given in terms of specific Lorentz components
of the single--nucleon tensor ${\cal W}_{m m'}^{\mu\nu}$, as discussed in
Refs.~[3,16].
The hadronic response functions entering in the coincidence cross section
can now be written in terms of the quantities defined above:
$$R^K=(2\pi)^3\sum_{m m'}{\cal R}^{K}_{m m'}n_{m m'}({\bf p},\Omega^*)
\ \ .\eqno(12)$$
One should note that in obtaining the single--nucleon responses (see
Refs.~[3,4]) the $z$--component is usually eliminated in favor of the
zero--component using current conservation (the $z$--axis is along ${\bf q}$).
One could also proceed the other way
and eliminate the charge components in favor of the longitudinal components.
Both are completely equivalent for an on--shell single--nucleon (conserved)
electromagnetic current. However,
in the case of off--shell nucleons
various prescriptions are usually employed that can
lead to different results when the longitudinal or charge components
(or neither) are eliminated via the property of current conservation.
Detailed discussions of this problem have already been
presented in Ref.~[3] and, drawing on those, in next section we only
summarize the main results and general
properties of the single--nucleon response functions that are relevant
for this work.
As discussed in Sect.~1, our aim in this paper is much more concerned
with the spin--dependent nuclear problem contained in the polarized spectral
function and its application to the case of deformed systems (in particular
$^{21}$Ne). However, the knowledge of the symmetries displayed by both types of
responses (single--nucleon and nuclear) will allow us to discuss in a very
straightforward way the general properties and symmetries introduced
by the polarization of the target and
incident electron.

\goodbreak\bigskip
\noindent{\bf 2.3\quad Half--off--shell Single--nucleon Responses}
\medskip\nobreak
In the Impulse Approximation one has to deal with the half--off--shell
$\gamma NN$ vertex. At present, there is not yet any rigorous approach to
treat the off--shellness property of the bound single--nucleon
current.\footnote{$^{\dagger}$}{For a review of this subject, see
Refs.~[18--22].}
Therefore, it has become common practice to use specific off--shell
extrapolations of the on--shell vertex.$^{[2,23,24]}$ The most frequently
used in the analysis of experimental data are the ones introduced
by de~Forest.$^{[4]}$ They are constructed via the following three steps:

a) Treat the spinors as free;

b) Employ the so--called $CC1$ and $CC2$ forms for the vertex operator;

c) Impose (or not) current conservation by elimination of the longitudinal
contributions
in favor of the charge contributions (or {\it vice versa}).

Following the work of de~Forest for the totally unpolarized
situation,$^{[4]}$ some of us (J.A.C., T.W.D. and G.I.P.)
generalized those previous studies to
the case in which the target and incident electrons are polarized. We
showed that, given a specific de~Forest--type prescription for the off--shell
vertex $\Gamma^{\mu}$, one can express the spin--dependent single--nucleon
tensor in terms of two tensors ${\cal S}^{\mu \nu}$ and
${\cal A}^{\mu \nu}_{m m'}$, where ${\cal S}^{\mu \nu}$ is real and
symmetric under the interchange $\mu \leftrightarrow \nu$,
whereas ${\cal A}^{\mu \nu}_{m m'}$ is antisymmetric under the interchange
$\mu \leftrightarrow \nu$, and has
real diagonal terms and in general complex off--diagonal terms.
The cross sections resulting from both recipes for the single--nucleon
tensor when current conservation is not imposed (denoted by $NCC1$ and
$NCC2$) were compared with the so--called $CC1/CC2$ cross sections
obtained by enforcing current conservation through elimination of the
$\mu=3$ or $\mu=0$
components of the current matrix elements by means of the relation
$qJ^{(3)}=\omega J^{(o)}$ (denoted by $CC1^{(0)}$/$CC2^{(0)}$ or
$CC1^{(3)}$/$CC2^{(3)}$, respectively).

By enforcing current conservation
and using the expansion of the responses in terms of symmetric and
antisymmetric parts,$^{[3]}$ one sees that the
electron--unpolarized single--nucleon response functions ${\cal R}^K_{mm'}$,
($K=L,T,TL$ and $TT$) are diagonal in the spin indices and take
the same value for both components ($m=m'=+,-$).
Hence, for simplicity we will denote them by
${\cal R}^K$. One can also show that
${\cal R}^{L}$ and ${\cal R}^T$ are independent of $\phi_N$,
while ${\cal R}^{TL}$ and ${\cal R}^{TT}$ are proportional to
$\cos\phi_N$ and $\cos 2\phi_N$, respectively.
The explicit $CC1^{(0)}$ and $CC2^{(0)}$
expressions of these responses can be found in Ref.~[3].
It is interesting to remark that the longitudinal response
${\cal R}^{L}$ is the
same for these two prescriptions, while this is not the case for
$CC1^{(3)}$ and $CC2^{(3)}$ prescriptions. This indicates that the
ambiguities introduced by the possible choices of $\Gamma^{\mu}$ for the
off--shell case are minimized when current conservation is imposed to
eliminate $J^{(3)}$ in favor of $J^{(0)}$. Thus, we have chosen this
prescription
($CC1^{(0)}$) for the calculations presented in this work.

In the case of the electron--polarized single--nucleon responses
${\cal R}^{K'}_{m m'}$, ($K'=T',TL'$), one can show that for any
off--shell vertex operator $\Gamma^{\mu}$, the
$\phi_N$--dependence can be written as follows:$^{[3]}$
$$\eqalignno{ {\cal R}^{T'}_l &= {1\over 2}({\cal R}^{T'}_{++}
-{\cal R}^{T'}_{--})= A &(13a)\cr
{\cal R}^{T'}_s  &= Re({\cal R}^{T'}_{+-})=
B\cos\phi_N &(13b)\cr
{\cal R}^{T'}_n  &= Im({\cal R}^{T'}_{+-})=
B\sin\phi_N &(13c)\cr
 {\cal R}^{TL'}_l &= {1\over 2}({\cal R}^{TL'}_{++}-{\cal R}^{TL'}_{--})=
C\cos\phi_N &(13d)\cr
{\cal R}^{TL'}_s  &= Re({\cal R}^{TL'}_{+-})=
{1\over 2}(D-E) + {1\over 2}(D+E)\cos 2\phi_N &(13e)\cr
{\cal R}^{TL'}_n  &= Im({\cal R}^{TL'}_{+-})=
{1\over 2}(D+E)\sin 2\phi_N \ \ , &(13f)\cr}$$
where $A$, $B$, $C$, $D$, $E$ are functions of $\theta_N$.
In particular, functions $A$ and $D\pm E$ are
symmetric in $\theta_N$, while $B$ and $C$ are antisymmetric.  For parallel
kinematics, where $\theta_N\rightarrow 0$, we find that $A$ and $D-E$ can be
nonzero whereas $B$, $C$ and $D+E$ must vanish.
Note that a relationship exists involving
two of the $T'$ responses: $\quad {\cal R}^{T'}_n /{\cal R}^{T'}_s =
\tan\phi_N$.
In the particular cases of the prescriptions $CC1^{(0)}$ and
$CC2^{(0)}$, the explicit expressions for the terms
$A,B,C,D,E$ can be found in Ref.~[3].

\goodbreak\bigskip
\noindent{\bf 3.\quad POLARIZED SPECTRAL FUNCTION AND DEFORMED NUCLEI}
\medskip\nobreak
In this section we develop a general expression for the polarized spectral
function spin matrix which enters in the
analysis of the reaction $\vec{A}(\vec{e},e'N)B$ in the PWIA. As previously
stated, we will restrict our attention to some simplified situations:
in particular, we will always consider the final nucleus to be left in
a bound state and assume parity conservation.
The nuclei involved in the process will be described by the rotational
model of Bohr and Mottelson using Nilsson and Hartree--Fock single--particle
wave functions. General properties and symmetries will be discussed in
Sect.~3.1.

The general expression for the
spin--dependent spectral function is
$$\eqalign{
&S_{mm'}({\bf p},E_m,\Omega^*)=\cr
&=\sum_{M_A}p(M_A)\sum_B
<B|a_{{\bf p}m'}|A>^{*}<B|a_{{\bf p}m}|A>
\delta (E_m+\epsilon_B-\epsilon_A^o)=\cr
&=\sum_{M_A}p(M_A)\sum_B
<A|a^+_{{\bf p}m'}|B><A|a^+_{{\bf p}m}|B>^{*}
\delta (E_m+\epsilon_B-\epsilon_A^o)\ \ , \cr}
\eqno(14)$$
where $|B>$ and $|A>$ represent the final and initial nuclear states,
respectively, and $a_{{\bf p}m}$, ($a^+_{{\bf p}m}$),
are annihilation (creation) operators which
destroy (create) a nucleon with momentum ${\bf p}$ and spin projection
$m$. Since we are considering the case of discrete nuclear states
we have good quantum numbers $J_A$ and $J_B$, respectively, as well as
parities $\pi_A$ and $\pi_B$. We will refer all
the quantities to the system defined by the axes $x,y,z$ (see Fig.~2).
This means that the final unpolarized nucleus will be characterized by
state vectors $|J_B M_B>$ defined with respect to the $z$--axis.
On the other hand,
the target nucleus is polarized, {\it i.e.,\/} the target is prepared with
magnetic substates $|J_A M_A>$ in the direction ${\bf P}^*$ populated in
a non--uniform manner with
probabilities $p(M_A)$. Thus, the target nuclear states are
quantized with respect to a fixed quantization axis
specified by the spherical coordinates $\Omega^*$.
We therefore have,
$$|J_AM_A>=
\sum_{M'_A}{\cal D}_{M_AM'_A}^{*J_A}(\Omega^*)|J_AM'_A>\ \ , \eqno(15)$$
where the eigenstates $|J_AM'_A>$ are referred to
the system with axes of quantization
along ${\bf q}$. We follow the convention of Edmonds$^{[25]}$ for
rotation matrices.\footnote{$^{\dagger}$}{Note that the convention of Brink and
Satchler (BS) is related to that of Edmonds (Ed) through the relation
$\left[{\cal D}_{MK}^{*I}\right]_{BS}
=\left[{\cal D}_{KM}^I \right]_{Ed}$.}

Expanding the single--nucleon creation (annihilation) operators over a
basis of irreducible tensor operators $a^+_{\ell jm_j}(p)$
($\tilde{a}_{\ell jm_j}(p)$)
and after some algebra (see
Appendix A for details), the polarized spectral function can be written
in terms of different tensor polarization components as
$$ S_{mm'}({\bf p},E_m,\Omega^*)=
\sum_I S_{mm'}^{(I)}({\bf p},E_m,\Omega^*)\ \ , \eqno(16)$$
where $I$ denotes the polarization rank ($I=0$
corresponds to the complete unpolarized case, {\it i.e.,\/}
target and incident electrons unpolarized).
Each of the tensor
polarization contributions in the spectral function
for parity--conserving electron scattering are given finally by
$$\eqalign{&S_{mm'}^{(I)}({\bf p},E_m,\Omega^*)
=(-1)^{m-1/2}f_I^{J_A} \sum_{B}
\sum_{\ell\ell'}\sum_{jj'}\sum_{LK} \sum_{M}
(-1)^{J_A+J_B+j+\ell'}
[j][j'][\ell][\ell'][L][K]^2\cr
&\times C^{*J_A J_B}_{\ell j}(p)C^{J_A J_B}_{\ell'j'}(p)
Y_{I}^{-M}(\Omega^*)Y_{L}^{-H}(\Omega)
\left( \matrix{\ell&\ell'&L\cr0&0&0\cr}\right)
\left(\matrix{ L&I&K\cr H&M&N\cr}\right)\cr
&\times
\left(\matrix{K&1/2&1/2\cr N&m&-m'\cr}\right)
\left\{\matrix{J_A&J_A&I\cr j&j'&J_B\cr}\right\}
\left\{ \matrix{L&I&K\cr\ell&j&1/2\cr\ell'&j'&1/2\cr}\right\}
\delta(E_m+\epsilon_{B}-\epsilon_{A}^o)\ \ , \cr}\eqno(17)$$
where $f_I^{J_A}$ are the spherical Fano statistical tensors as given in
Eq.~(A.7) and
$C^{*J_A J_B}_{\ell j}(p)$ $\left[ C^{J_A J_B}_{\ell' j'}(p)\right]$
the reduced nuclear matrix elements (see Appendix A).
Note that all of the angular dependence is contained in the two spherical
harmonics, $Y_I^{-M}(\Omega^*)$ and $Y_L^{-H}(\Omega)$ with
$\Omega\equiv(\theta,
\phi)$, the angles defining the direction of the struck nucleon in the
laboratory frame.

A sum over the single--particle quantum numbers $\ell,\ell'$
($j,j'$) is involved in the expression for the
spectral function and introduces interferences between different
single--nucleon orbitals. This arises only when polarization
degrees of freedom are taken into account$^{[26]}$.
In the complete unpolarized case
(see Appendix A), only the trace of $S_{mm'}$ enters; in other words the
spectral function is reduced to
$$ S_{mm'}(p,E_m)=S(p,E_m)\delta_{mm'}=
{1\over 8\pi} {1\over [J_A]^2}\sum_B \sum_{\ell j}
|C^{J_A J_B}_{\ell j}(p)|^2\delta(E_m-\epsilon_{B}+\epsilon_{A}^o)
\delta_{mm'} \ \ ,\eqno(18)$$
and therefore, the interference terms do not appear.
These effects will be studied in detail in a forthcoming work. In the present
paper we are interested in studying
the general behaviour of the differential
cross sections as well as of the
hadronic response functions and polarization ratios as functions of the
bound nucleon momentum, the target polarization direction and
the deformation of the target.

\goodbreak\bigskip
\noindent{\bf 3.1\quad Properties and Symmetries of the Spectral Function}
\medskip\nobreak
A property that emerges from Eq. (10b) is that
$$S_{m m'}^{(I)}({\bf p},E_m,\Omega^*)=
S_{m' m}^{(I)*}({\bf p},E_m,\Omega^*)\ \ . \eqno(19)$$
On the other hand, using parity conservation
it is straightforward to show that,
$$S_{-m -m'}^{(I)}({\bf p},E_m,\Omega^*)=
(-1)^{I+1-m-m'}
S_{mm'}^{*(I)}({\bf p},E_m,\Omega^*)\ \ . \eqno(20)$$
Combining both relations it is easy to see that the
off--diagonal components of the spectral function are zero for
$I=$ even.

For convenience, we introduce the components
$S_0$, $S_{l}$, $S_s$ and $S_n$ which are defined to be real
(for clarity, here we do not specify
dependence of the spectral function on momentum, energy and angular
variables),
$$\eqalignno{S_0 &\equiv S_{++}+S_{--} &(21a)\cr
S_{l} &\equiv S_{++}-S_{--} &(21b)\cr
S_s &\equiv S_{+-}+S_{-+} =2Re(S_{+-}) &(21c)\cr
S_n &\equiv -i(S_{+-}-S_{-+}) =2Im(S_{+-})\ \ . &(21d)\cr}$$
Using the properties given by Eqs.~(19,20), it follows
that $S_0$ only has contributions from even $I$--values, whereas
$S_l$, $S_s$ and $S_n$ only have contributions from odd $I$--values.
One can also see that
$$\eqalignno{S_0 &=
\sum_{I={\hbox{even}}} S_0^{(I)}({\bf p},E_m,\Omega^*)=
2\sum_{I={\hbox{even}}}
S^{(I)}_{mm}({\bf p},E_m,\Omega^*) &(22a)\cr
S_{l} &=
\sum_{I={\hbox{odd}}} S_l^{(I)}({\bf p},E_m,\Omega^*)=
2(-1)^{1/2-m}\sum_{I={\hbox{odd}}} S^{(I)}_{mm}
({\bf p},E_m,\Omega^*) &(22b)\cr
S_s &=
\sum_{I={\hbox{odd}}} S_s^{(I)}({\bf p},E_m,\Omega^*)=
2\sum_{I={\hbox{odd}}} Re\left[S^{(I)}_{m-m}
({\bf p},E_m,\Omega^*)\right] &(22c)\cr
S_n &=
\sum_{I={\hbox{odd}}} S_n^{(I)}({\bf p},E_m,\Omega^*)=
2(-1)^{1/2-m} \sum_{I={\hbox{odd}}}
Im\left[S^{(I)}_{m-m}
({\bf p},E_m,\Omega^*)\right]\ \ . &(22d)\cr}$$
With this new notation, one should remember that the 0--component is the
only one that contributes
in the electron--unpolarized responses (it may include contributions from
target polarization, $I\geq 2$). The other three components $l$, $s$ and $n$
enter in the two electron--polarized responses
(electrons and target polarized).

The electron--unpolarized (helicity sum)
and electron--polarized (helicity difference) cross sections (see Sect.~2.1)
which enter in the definition of the global asymmetry
$A=\Delta/\Sigma$ can now simply be written
$$\eqalign{\Sigma=&
{p_N M_N M_B\over M_A}f_{rec}^{-1}
\sigma^{eN}_0 n_0({\bf p},\Omega^*)\cr
\Delta=&
{p_N M_N M_B\over M_A}f_{rec}^{-1}
\left[\sigma^{eN}_l n_l({\bf p},\Omega^*)+
\sigma^{eN}_s n_s({\bf p},\Omega^*)-
\sigma^{eN}_n n_n({\bf p},\Omega^*)\right]\ \ , \cr}\eqno(23)$$
where $\sigma^{eN}_i$, ($i=0,l,s,n$) are the components for the
``off--shell electron--nucleon cross section'' defined in terms of
$\sigma_{m m'}^{eN}$ in analogy to Eqs.~(21) but divided by a factor 2
(see Ref.~[3]), and
$n_i({\bf p},\Omega^*)$ are the different
components of the spin--dependent momentum distribution (Eq.~(9))
defined in the same way as the components of the polarized spectral function.
The hadronic response functions (Eq.~(12)) are then given by
$$R^K=(2\pi)^{3}
{\cal R}^K n_0({\bf p},\Omega^*)\eqno(24a)$$
$$R^{K'}=(2\pi)^{3} \left[
{\cal R}^{K'}_l n_l({\bf p},\Omega^*)+
{\cal R}^{K'}_s n_s({\bf p},\Omega^*)-
{\cal R}^{K'}_n n_n({\bf p},\Omega^*)\right]\ \ ,
\eqno(24b)$$
where $K=L,T,TL,TT$ and $K'=T',TL'$.
The single--nucleon response functions, ${\cal R}^{(K/K')}_i$
are specified in the way discussed in Sect.~2.3 (see Ref.~[3] for details).

In connection with the angular dependence in the polarized spectral function,
one should note that the whole dependence in Eq.~(17)
is given through the two spherical harmonics,
$Y_I^{-M}(\Omega^*)$ and $Y_L^{-H}(\Omega)$, with
$\Omega\equiv\{\theta,\phi\}$ and
$\Omega^*\equiv\{\theta^*,\phi^*\}$ being the
angles defining the direction of the struck
nucleon momentum ${\bf p}$ and
the direction of the target polarization ${\bf P^*}$, respectively.
{}From Eqs.~(17,21) the different components of the spin--dependent
momentum distribution for a fixed $J_B$--value can be written
as follows:

a) The component entering in the electron--unpolarized cross section is
given by
$$n^{J_B}_0=
\sum_{I={\hbox{even}}}[I]f_I^{J_A} {\cal J}(J_A,J_B,I;p)P_I(\cos\xi)
\ \ , \eqno(25)$$
where ${\cal J}(J_A,J_B,I,p)$ contains
all of the dependence on the model used in the
evaluation of the nuclear wave functions. As noted above, it depends on the
total angular momenta of the target and residual nucleus, on the rank of the
polarization tensor and on the magnitude of
the struck--nucleon momentum. Its explicit expression in terms of the
nuclear reduced matrix elements is given by
$$\eqalign{{\cal J}(J_A,J_B,I;p)=&
{(-1)^{J_A+J_B+1/2}\over 4\pi}\sum_{\ell \ell'}\sum_{jj'}
(-1)^{2j}[j][j'][\ell][\ell']C^{*J_A J_B}_{\ell j}(p)
C^{J_A J_B}_{\ell'j'}(p)\cr
&\times\left( \matrix{\ell&\ell'&I\cr0&0&0\cr}\right)
\left\{\matrix{J_A&J_A&I\cr j&j'&J_B\cr}\right\}
\left\{\matrix{j&j'&I\cr \ell'&\ell&1/2\cr}\right\}\ \ .
\cr}\eqno(26)$$
The angle $\xi$ entering in the Legendre polynomial $P_I(\cos\xi)$
is the relative angle between the direction of the target polarization
and the momentum of the bound nucleon. It is given through the relation
$$\cos\xi=\cos\theta^* \cos\theta +\sin\theta^*\sin\theta\cos\Delta \phi
\ \ , \eqno(27)$$
where
$$\Delta\phi=\phi^*-\phi_N=\phi^*-\phi \eqno(28)$$
is the angle between the planes (${\bf q}, {\bf P}^*$) and
(${\bf q},{\bf p}_N$) [or equivalently (${\bf q},{\bf p}$)].

b) The three components that enter in the electron--polarized cross section
are given by
$$\eqalign{n_l^{J_B}
=&\sum_{I={\hbox{odd}}}[I]f_I^{J_A}\sum_{L=I-1}^{I+1}
{\cal K}(J_A,J_B,I,L;p)\cr
&\times
\sum_{M=-I}^I {\cal X}(I,L,M,\theta)
\sqrt{{(I+M)!\over (I-M)!}} P_I^{-M}
(\cos\theta^*)\cos M\Delta\phi
\cr}\eqno(29a)$$
\goodbreak\medskip
$$\eqalign{n_{s}^{J_B}=&\sum_{I={\hbox{odd}}}[I]f_I^{J_A}\sum_{L=I-1}^{I+1}
{\cal K}(J_A,J_B,I,L;p)\sum_{M=-I}^I {\cal Z}(I,L,M;\theta)
\cr
&\times \sqrt{{(I+M)!\over (I-M)!}} P_I^{-M}
(\cos\theta^*)
\cos (M\Delta\phi+\phi_N)
\cr}\eqno(29b)$$
\goodbreak\medskip
$$\eqalign{n_{n}^{J_B}=&-\sum_{I={\hbox{odd}}}[I]f_I^{J_A}\sum_{L=I-1}^{I+1}
{\cal K}(J_A,J_B,I,L;p)\sum_{M=-I}^I {\cal Z}(I,L,M;\theta)
\cr
&\times  \sqrt{{(I+M)!\over (I-M)!}} P_I^{-M}
(\cos\theta^*)
\sin (M\Delta\phi+\phi_N)\ \ .
\cr}\eqno(29c)$$
The common function ${\cal K}(J_A,J_B,I,L;p)$, with a form similar to
${\cal J}(J_A,J_B,I;p)$, contains all of the nuclear structure dependence.
Its explicit expression is
$$\eqalign{{\cal K}(J_A,J_B,I,L;p)=&
{\sqrt{3}\over 2\pi}(-1)^{J_A+J_B}[L]^2
\sum_{\ell \ell'}\sum_{jj'}(-1)^{j+\ell}[j][j'][\ell][\ell']
C^{*J_A J_B}_{\ell j}(p)C^{J_A J_B}_{\ell'j'}(p)\cr
&\times \left( \matrix{\ell&\ell'&L\cr0&0&0\cr}\right)
\left\{\matrix{J_A&J_A&I\cr j&j'&J_B\cr}\right\}
\left\{ \matrix{L&I&1\cr\ell&j&1/2\cr\ell'&j'&1/2\cr}\right\}\ \ .
\cr}\eqno(30)$$
The functions
${\cal X}(I,L,M;\theta)$ and ${\cal Z}(I,L,M;\theta)$
are given by
$$\eqalign{{\cal X}(I,L,M;\theta)=
{1\over \sqrt{2}}
\sqrt{{(L-M)!\over (L+M)!}}
\left(\matrix{ L&I&1\cr -M&M&0\cr}\right)
P_L^{M}(\cos\theta)
\cr}\eqno(31a)$$
$$\eqalign{{\cal Z}(I,L,M;\theta)=
\sqrt{{(L-M+1)!\over (L+M-1)!}}
\left(\matrix{ L&I&1\cr 1-M&M&-1\cr}\right)
P_L^{M-1}(\cos\theta)\ \ .
\cr}\eqno(31b)$$
Note that the relation, ${\cal X}(I,L,-M;\theta)=(-1)^M{\cal X}(I,L,M;\theta)$
holds. With regards to the dependence on the azimuthal angles, $n_o^{J_B}$ and
$n_l^{J_B}$
depend only on $\Delta\phi$, while $n_s^{J_B}$ and $n_n^{J_B}$
depend on both $\Delta\phi$ and $\phi_N$.

\goodbreak\bigskip
\noindent{\bf 3.2\quad Application to Deformed Nuclear Systems}
\medskip\nobreak
In the general expression for the polarized spectral function, one should note
that the dependence on the nuclear model is contained in the
reduced nuclear matrix elements $C_{\ell j}^{J_A J_B}(p)$.
As previously stated, we focus on the case of rotational
nuclei and use the
factorization approximation of Bohr and Mottelson$^{[6]}$.
The wave function for axially--symmetric
deformed nuclei in the laboratory system is written
in terms of the relative orientation of the body--fixed system and the
intrinsic wave function. For a given $K_A$--band in the $A$--body nucleus with
intrinsic (Slater determinant or BCS) wave function $\Phi_{K_A}$, the
nuclear wave function is given by$^{[6]}$
$$\eqalign{|\Psi_A J_A M_A\rangle \equiv |J_A K_A M_A\rangle &=
\left( {2J_A+1 \over 16\pi^2(1+\delta_{K_A,0})}\right)^{1/2}\cr
&\times \left[{\cal D}^{J_A}_{K_A M_A}\Phi_{K_A} + (-1)^{J_A+K_A}
{\cal D}^{J_A}_{-K_A M_A}\Phi_{\overline{K}_A}\right]\cr}\eqno(32)$$
and a similar expression holds
for the wave function corresponding to the residual nucleus $B$.
With these nuclear wave functions,
the reduced nuclear matrix elements that result are given by
$$\eqalignno{C^{J_A J_B}_{\ell j}(p)=&{(-1)^{J_A-K_A} [J_A] [J_B]
\over \sqrt{(1+\delta_{K_A,0})(1+\delta_{K_B,0})}}\sum_{\mu}
\biggl[\left(\matrix{J_A&j&J_B\cr-K_A& \mu&K_B \cr}\right)
\langle \Phi_{K_A}|a^+_{\ell j\mu}(p)|\Phi_{K_B}\rangle +\cr
&(-1)^{J_B+K_B}
\left(\matrix{J_A&j&J_B\cr -K_A& \mu&-K_B \cr}\right)
\langle \Phi_{K_A}|a^+_{\ell j\mu}(p)|\Phi_{\overline{K}_B}
\rangle\biggr]\ \ , &(33)\cr}$$
where $a^+_{\ell j\mu}(p)$ is the creation operator for a nucleon with
momentum $p$ and
quantum numbers $\ell ,j,\mu$. The process considered involves a polarized
odd--A target and therefore $J_A$ and $K_A$ take on half--integer values.
We only consider
transitions from the ground state of the target nucleus
to states in the residual
nucleus within the ground--state band, $K_B=0$. In such a case, the reduced
matrix elements are then just given by
$$C^{J_A J_B}_{\ell j}(p)=\sqrt{2}(-1)^{J_A-K_A}[J_A][J_B]
\left(\matrix{J_A&j&J_B\cr -K_A& K_A&0 \cr}\right)
\langle \Phi_{K_A}|a^+_{\ell jK_A}(p)|0\rangle\ \ , \eqno(34)$$
with
$$\langle 0|a_{\ell jK_A}(p)|\Phi_{K_A}\rangle=\sqrt{2
v^2_{K_A}}\sum_{n}c^{K_A}_{n\ell j}\psi_{n\ell j}(p)\ \ .\eqno(35)$$
Here $v^2_{K_A}$ is the probability for the orbit $K_A$ to be occupied in the
target nucleus. Calculations for the intrinsic state of the odd--$A$ nucleus
are done in the pair--filling approximation, fixing the $K_A$--orbital of
the odd--nucleon, in which case $2v_{K_A}^2=1$. Finally,
$c_{n\ell j}^{K_A}$ are the amplitudes of the single--particle deformed
state $|\Phi_{K_A}\rangle$ in the spherical basis and $\psi_{n\ell j}(p)$ is
the Fourier transform of the $n l j$ radial wave function in such a basis.

\goodbreak\bigskip
\noindent{\bf 4.\quad RESPONSE FUNCTIONS AND POLARIZATION RATIOS}
\medskip\nobreak
In this section we obtain
the final expressions for the different hadronic response
functions and asymmetries and/or polarization ratios.
The differential cross section (Eq.~(1)) can be written in the following way
$${d\sigma^h \over d\epsilon' d\Omega_e d\Omega_N}=
\left[{d\sigma \over d\epsilon' d\Omega_e d\Omega_N}\right]_{0}
\left[1+{\cal P}_{\Sigma}+h{\cal P}_{\Delta}\right]\ \ , \eqno(36)$$
where $\left[{d\sigma \over d\epsilon' d\Omega_e d\Omega_N}\right]_{0}$
is the differential cross section for target and electron unpolarized.
In terms of the single--nucleon
responses (${\cal R}^K$) and reduced nuclear matrix elements ($C_{\ell j}
^{J_A J_B}(p)$), it is given by
$$\eqalign{
\left[{d\sigma \over d\epsilon' d\Omega_e d\Omega_N}\right]_{0}&=
{p_N M_N M_B \over (2\pi)^3 M_A}\sigma_{Mott}f_{rec}^{-1}[R]_{0}\cr
&={p_N M_N M_B \over 4\pi M_A}\sigma_{Mott}f_{rec}^{-1}
{[\sum_{B}\sum_{\ell j}|C_{\ell j}^{J_A J_B}(p)|^2] \over [J_A]^2}
\sum_K v_K {\cal R}^K\ \ , \cr}\eqno(37)$$
where the sum extends over $K=L,T,TL,TT$.
The terms ${\cal P}_{\Sigma}$ and ${\cal P}_{\Delta}$ are
the polarization ratios. The fraction, ${\cal P}_{\Delta}/
({\cal P}_{\Sigma}+1)$ measures the relationship between
the helicity--difference (electron--polarized) and the helicity--sum
(electron--unpolarized) cross sections.
The term ${\cal P}_{\Sigma}$
gives the target asymmetry when the electron beam is unpolarized,
whereas
${\cal P}_{\Delta}$ takes into account as well the contribution coming from
the polarization of the incident electrons.

Let us first discuss some general properties of ${\cal P}_{\Sigma}$.
It is important to note that ${\cal P}_{\Sigma}$ is independent of the
single--nucleon cross section ($\sigma^{eN}_{m m'}$) (Eq.~(23)).
Therefore, it does not
depend on the single--nucleon current and is free from the ambiguities
introduced by the various choices ($CC1^{(0)},CC2^{(0)},CC1^{(3)},CC2^{(3)},
NCC1$ and $NCC2$) discussed in Ref.~[3]. In this respect, ${\cal P}_{\Sigma}$
may be an ideal tool for studying nuclear structure, in particular the
spin--dependent momentum distribution of bound nucleons.

Using the expression given in the previous section for the 0--component
of the spectral function (Eqs.~(25,26)) we can obtain explicit expressions
for the target polarization ratio. This will allow us to study in a very simple
way the behaviour of ${\cal P}_{\Sigma}$ for the nucleus of interest here
($^{21}$Ne) at different kinematics. The
target polarization ratio for a fixed--$J_B$ state
can be written
$${\cal P}_{\Sigma}^{J_B}={4\pi[J_A]^2\over \left[\sum_{\ell j}
|C^{J_A J_B}_{\ell j}(p)|^2
\right]}
\sum_{I\geq 2,\hbox{even}}
[I]f_I^{J_A}{\cal J}(J_A,J_B,I;p) P_I(\cos\xi)\ \ ,\eqno(38)$$
where ${\cal J}(J_A,J_B,I;p)$ is defined in Eq.~(26) and $f_I^{J_A}$ is
the Fano statistical tensor given in Eq.~(A.7).
Note that ${\cal P}^{J_B}_{\Sigma}$ depends in general on the
nuclear model. $C_{\ell j}^{J_A J_B}(p)$ is
the $\ell ,j$--component of the bound nucleon that is knocked--out in the
transition from $J_A$ to $J_B$ (see Eqs.~(33--35)). The only angular dependence
of ${\cal P}_{\Sigma}$ is given through $\xi$, the relative angle between
the direction of the polarization vector and the momentum of the bound nucleon
(see Eq.~(27)). The particular case in which the final
nucleus is considered to be in its ground state, {\it i.e.\/}
$J_B=0^+$, is specially
simple: there the values of the single--nucleon angular momenta
are fixed and $j$ and $\ell$ take on single values
($j=j'=J_A$ and $\ell=\ell'$). The target polarization ratio is then reduced to
$$
{\cal P}_{\Sigma}^{J_B=0}=
(-1)^{J_A+1/2}[J_A]^2[\ell]^2\sum_{I\geq 2,\hbox{even}}[I]f_I^{J_A}
\left ( \matrix{\ell &\ell &I \cr 0&0&0\cr}\right)
\left \{ \matrix{J_A &J_A &I \cr \ell&\ell&1/2\cr}\right\}
P_I(\cos\xi)\ \ , \eqno(39)$$
where the value of $\ell=J_A\pm 1/2$ is fixed by the parity and angular
momentum of the target nucleus. Note that
${\cal P}_{\Sigma}^{J_B=0}$ is independent of the model used
for the description of the single--particle
wave functions.

The structure of ${\cal P}_{\Delta}$ (electron--target polarization ratio) is
much more complex due to the new degrees of freedom introduced by the
polarization of the incident electrons. In this situation one needs
to evaluate three components of the spectral function ($S_l$,
$S_s$, $S_n$), as well as the three components of the single--nucleon
responses (${\cal R}^{K'}_l, {\cal R}^{K'}_s, {\cal R}^{K'}_n$).
In order to simplify the
discussion, we can decompose the electron--target polarization
${\cal P}_{\Delta}$ in the following way:
$${\cal P}_{\Delta}=\sum_{K'=T',TL'}v_{K'}{\cal P}_{\Delta}^{K'}\ \ ,
\eqno(40a)$$
where $v_{K'}$ are the kinematical factors$^{[16]}$ and
${\cal P}_{\Delta}^{K'}$ are polarization ratios given by
$${\cal P}_{\Delta}^{K'}={R^{K'}\over [R]_{0}}
\eqno(40b)$$
with $[R]_{0}$ as given in Eq.~(37).
Using the explicit expressions for the various single--nucleon
responses and for the components of the spectral function
given in Sect.~2.3 and 3, respectively, one finds the following expressions
for the two hadronic response functions involved:
$$\eqalign{
R^{T'}=&
(2\pi)^3 \sum_{I={\hbox{odd}}}[I]f_I^{J_A}\sum_{L=I-1}^{I+1}{\cal K}
(J_A,J_B,I,L;p)
\sum_{M=-I}^I \sqrt{{(I+M)!\over(I-M)!}} P_I^{-M}(\cos\theta^*)
\cr
&\times \biggl [A{\cal X}(I,L,M;\theta)+B{\cal Z}(I,L,M;\theta)
\biggr ]
\cos M\Delta\phi
\cr}\eqno(41a)$$
$$\eqalign{
R^{TL'}=&
(2\pi)^3 \sum_{I={\hbox{odd}}}[I]f_I^{J_A}\sum_{L=I-1}^{I+1}{\cal K}
(J_A,J_B,I,L;p)
\sum_{M=-I}^I \sqrt{{(I+M)!\over (I-M)!}} P_I^{-M}(\cos\theta^*)
\cr
&\times
\biggl[ \left[ C{\cal X}(I,L,M;\theta)
+D{\cal Z}(I,L,M;\theta) \right]\cos M\Delta\phi\cos\phi_N
\cr
&\qquad -E{\cal Z}(I,L,M;\theta)\sin M\Delta\phi \sin\phi_N
\biggr]\ \ ,
\cr}\eqno(41b)$$
where again the expressions refer to a fixed $J_B$--value of
the residual nucleus.
The terms, $A,B,C,D,E$ were introduced in
Sect.~2.3. The function
${\cal K}(J_A,J_B,I,L;p)$ that contains the dependence on
the $C_{\ell j}^{J_A J_B}(p)$ nuclear amplitudes, as well as the functions
${\cal X}(I,L,M;\theta)$ and ${\cal Z}(I,L,M;\theta)$
are given in Eqs.~(30,31).

Some very general properties for these electron--polarized
hadronic responses now become evident.
First, we note that the dependence on the single--nucleon responses
cannot be factored--out, as it could in the case of the
electron--unpolarized response
functions. Therefore, ${\cal P}_{\Delta}$ depends on the choice of the
single--nucleon current. On the other hand, Eqs.~(41) show the
explicit dependence on the azimuthal angles $\Delta\phi$ and $\phi_N$ of
the $T'$ and $TL'$ response functions. These expressions allow us
to study some particular situations where such functions can or cannot
be probed. For example, it is known that for $N$--polarized nuclei and
co--planar kinematics both electron--polarized responses are zero and
no additional information is gained by polarizing the electron. This situation
corresponds to $\theta^*=\phi^*=90^o$ (nucleus polarized in the $y$--direction)
and $\phi_N=0^o$ (nucleon detected in the scattering plane). From Eqs.~(41)
it is easy to see that both responses $T'$ and $TL'$ are zero.
For $\phi_N=0^o$ only the terms proportional to $\cos M\Delta\phi$ may
survive. Since $\Delta\phi=-90^o$, only the terms with $|M|=$ even may
contribute, but the Legendre polynomials are zero for $I=$ odd, $|M|=$ even and
$\theta^*=90^o$. In Table~1 we summarize the situations in which one or
both response functions ($T', TL'$) are zero.

\goodbreak\bigskip
\noindent{\bf 5.\quad RESULTS FOR THE REACTION
${}^{21}{\overrightarrow{\hbox{Ne}}}
({\vec{\hbox{e}}},\hbox{e}'\hbox{n})^{20}$Ne }
\medskip
\noindent{\bf 5.1\quad The Nuclear Model}
\medskip\nobreak
In this section we present and discuss
the results obtained for the cross sections, response functions and
polarization ratios introduced in the previous sections for the case
of $^{21}{\overrightarrow{\hbox{Ne}}}$. As discussed in Sect.~3.2 we
consider transitions to
states in the residual nucleus within the ground--state band, $K_B=0$, and
then the ejected nucleon is the last bound neutron.
The selection of $^{21}$Ne as the particular nucleus for which to apply all of
the formalism developed above arises for several reasons.
Firstly, from an experimental point of view, being a noble gas
$^{21}{\overrightarrow{\hbox{Ne}}}$ is only
weakly--reacting chemically and hence does not tend to depolarize when
undergoing collisions with the target--container walls.
Secondly, it presents a well--established rotational
energy spectrum with well--defined bands,$^{[5]}$ that is, the experimental
data support the description of such nucleus within the rotational model
of Bohr and Mottelson.$^{[6]}$ Finally, its charge is not very large
($Z=10$) and one does not need to worry unduely about Coulomb distortion of
the electron.

The two nuclei involved in the scattering reaction
(target and residual system) are taken as deformed
nuclear systems with axial symmetry and the same mean field. The deformed
single--particle wave functions have been calculated by using (1) the
phenomenological potential of the Nilsson model and
(2) the more sophisticated self--consistent density--dependent potentials of
deformed Hartree--Fock (DDHF) calculations. Cross sections, response functions
and polarization ratios obtained
with both models are compared in next section. Here we summarize the main
results of the Nilsson model and DDHF calculations that are relevant for the
discussion in next sections.

Let us start by making some very general remarks in the case of the
Nilsson model.$^{[27]}$ In this model one
introduces a deformed harmonic oscillator potential with spin--orbit
coupling terms and axial symmetry. The parametrization of such a potential
for the description of $^{21}$Ne has been taken from the literature$^{[28]}$
and pairing correlations have been omitted, as is the standard procedure
for $s$--$d$ shell nuclei.
In the case of $^{21}$Ne, the ground state is $J_A^{\pi}=3/2^+$ and the
last unpaired neutron is in the state
$K^{\pi}=3/2^+$. Eight
major shells ($N$) have been used in the diagonalization of the Nilsson
hamiltonian and $N$--admixtures have been considered.
The equilibrium Nilsson deformation parameter for
$^{21}$Ne in its ground state is found to be $\delta_{eq}=0.31$

In the Hartree Fock model,
the calculations have been made using the McMaster version of the deformed
Hartree-Fock code$^{[29]}$ that follows closely the method of
Ref.~[30]. In all the cases 50 iterations have been enough
for reaching a good convergence. The effective two-nucleon interaction
from which the average one-body field is obtained has been chosen to be
the Skyrme-type interaction SKA (Ref.~[31]). Tables~I and II of
Ref.~[12] summarize
the theoretical results obtained for $^{21}$Ne with different Skyrme
interactions using a deformed H.O. basis with $N_0=10$, and their comparison
to experimental data.
We summarize in Table~2 the results obtained with SKA interaction for
r.m.s. radii
and quadrupole moments in ${\bf r}$--space and in ${\bf p}$--space. Also
shown in this table are the results obtained with the Nilsson model at the
equilibrium deformation ($\delta_{eq}=0.31$) in terms of $b$ (the harmonic
oscillator length parameter in fm units). In this table $Q_0^r$ represents
the standard quadrupole moment for protons ($\pi$) or neutrons ($\nu$), and
$Q_0^p$ is defined likewise in ${\bf p}$--space
$$\eqalignno{
Q_0^r = & \int d{\bf r} \rho({\bf r}) P_2(\Omega_r) &(42a)\cr
Q_0^p = & \int d{\bf p} n({\bf p}) P_2(\Omega_p) &(42b)\cr}
$$
where $\rho({\bf r})$ and $n({\bf p})$ are the intrinsic one--body densities in
${\bf r}$--space and in ${\bf p}$--space, respectively.

The deformation parameters in ${\bf r}$ and in ${\bf p}$--spaces are defined
following the standard convention$^{[6]}$
$$\eqalignno{
\beta^r=&\sqrt{{\pi\over 5}}{Q_0^r \over A\langle r^2\rangle} &(43a)\cr
\beta^p=&\sqrt{{\pi\over 5}}{Q_0^p \over A\langle p^2\rangle} \ \ . &(43b)\cr}
$$
%

Note that the $\beta$ parameters corresponding to densities in ${\bf
p}$--space are much smaller than those in ${\bf r}$--space in agreement
with the fact that at the equilibrium deformation the equalities
$\langle p_x^2\rangle =\langle p_y^2\rangle = \langle p_z^2\rangle$ are
approximately satisfied$^{[6,13,14]}$. It is however interesting to note
that in this odd--A nucleus the $\beta^p$ value is larger than in the
even--even nuclei discussed in previous papers$^{[13,14]}$. This is
understood from the fact that in the odd--A case the unpaired nucleon
tends to increase the anisotropy in ${\bf p}$--space, which is however
smaller than that in ${\bf r}$--space even for the unpaired
nucleon. This is illustrated in Figure~4 where we compare the $\ell
=0,2$ multipoles of the single--particle density for the unpaired
orbital in ${\bf r}$--space and in ${\bf p}$--space. In this figure the
single--particle monopole densities in {\bf r} and {\bf p}--spaces are
denoted by $\rho_0(r)$ and $n_0(p)$ respectively and the quadrupole
densities are denoted by $\rho_2(r)$ and $n_2(p)$. They have been
calculated from the expressions $$\eqalignno{
\rho({\bf r})=&\sum_{\ell} \rho_{\ell}(r)P_{\ell}(\Omega_r)  &(44a)\cr
n({\bf p})=&\sum_{\ell} n_{\ell}(p) P_{\ell}(\Omega_p) &(44b)\cr}$$
with $\rho({\bf r})$ ($n({\bf p})$) the intrinsic density in {\bf r}--space
({\bf p}--space) corresponding to the deformed orbital $K^{\pi}=3/2^+$,
occupied by the odd neutron averaged on spin.

The DDHF results (solid lines) are compared to the results of Nilsson
model with (dashed lines) and without (short--dashed lines) major shell
N--admixtures. It is interesting to note that the nice agreement between
DDHF and Nilsson ($\Delta$N=2) results is destroyed when N--admixtures
are neglected in the Nilsson model. As seen in the figure in the latter
case (Nilsson ($\Delta$N=0)) the monopole and quadrupole densities
differ not only quantitatively but also qualitatively from the
corresponding DDHF results.  This is most noticeable for the quadrupole
density in {\bf p}--space ($n_2(p)$) that becomes negative at $p\geq 1$
fm$^{-1}$ in both Nilsson ($\Delta$N=2) and DDHF results. This important
feature of $n_2(p)$ is necessary to satisfy the isotropy condition
$\beta^p\approx 0$ at equilibrium$^{[6,13,14]}$. This feature is lost
when $\Delta$N=2 admixtures are neglected in the Nilsson model, and this
is the reason why $\Delta$N=2 admixtures have to be included when
discussing momentum distributions.  In Table~3 we give the $\ell j$
contributions to the normalization of the odd--neutron wave
function. The tabulated $n_{\ell j}$ weights have been calculated as
described in Refs.~[13,14] (see in particular Eq.~(27) in Ref.~[14]). As
seen in this table the dominant angular momentum component of the
$K^{\pi}=3/2^+$ wave function is $d_{5/2}$. In turn $92\%$ of this
component corresponds to the N=2 shell. On the overall the total
contribution from higher N shells (N$>2$) to this state amounts to $\sim
10\%$. The amount of admixtures from higher N shells in the other
occupied single--particle states is of this same order. Even though the
admixtures from higher N shells are small, they play a crucial role in
getting the correct $\beta$--values in {\bf r} and {\bf p}--spaces
simultaneously.

In Figs.~5--8 we show representative results for the unpolarized and
polarized momentum distributions given by Eqs.~(25) and (29).  For
brevity only the transition between the $^{21}$Ne and $^{20}$Ne ground
states is considered here and the struck--nucleon momentum {\bf p} is
taken to lie in the $xz$--plane, so that the momentum distributions can
be displayed as surfaces with $p_x$ and $p_z$ as independent variables.
We begin with Figs.~5--7 in which the target nucleus
$^{21}{\overrightarrow{\hbox{Ne}}}$ is assumed to be polarized in the
$z$--direction (along {\bf q}). The electron--unpolarized momentum
distribution $n_0$ for this case is shown in Fig.~5.  It exhibits peaks
symmetrically placed along the $p_x$--axis and a nodal line lying along
the $p_z$--axis.  If the target is polarized in the $x$--direction
instead of the $z$--direction, then the same functional behaviour is
obtained except that $p_x\leftrightarrow p_z$.  The polarized momentum
distributions $n_l$ and $n_s$ (for target polarization in the
$z$--direction) are displayed in Figs.~6 and 7.  In these cases, if the
polarization is placed along the $x$--axis, then $n_l$ and $n_s$ are
interchanged and as before $p_x\leftrightarrow p_z$.  For target
polarization along either $z$-- or $x$--axes one has $n_n=0$.  Finally,
for target polarization placed in the $y$--direction $n_0=n_n$ is that
shown in Fig.~8, while $n_l=n_s=0$.

Thus, we see considerable richness in the momentum distributions for
polarized targets of the types accessible both with and without
polarized electrons.  For unpolarized targets, on the other hand, the
distributions must be averaged over all angles and consequently much of
this richness will be lost.  In that case, the momentum distribution
will depend only on the magnitude of {\bf p} (see Appendix A.1).
Effectively, orienting a polarized deformed nucleus means that the
nuclear matter is distributed asymmetrically in coordinate space (and
hence in momentum space as employed here).  When combined with the
polarized single--nucleon cross section to obtain the coincidence cross
section (as in the following results and discussions), one will see
reflections of these distributions: orienting the target in some
specific way one will find more cross section when the outgoing nucleon
emerges in some direction for {\bf p}$_N$ (and hence {\bf p}) than in
another.

\goodbreak\bigskip
\noindent{\bf 5.2\quad Electron Scattering and Polarization Kinematics}
\medskip\nobreak
Forward-- and backward--angle electron scattering situations
have been considered and in the following results are shown for both
co--planar ($\phi_N=0^o$) and out--of--plane
($\phi_N=90^o$) kinematics.
The incident electron beam and nuclear target are assumed to
be 100\% polarized, {\it i.e.,\/} the weighting factor $p(M_A)$
in the expression of the spectral function (Eq.~(14)) is given
by $p(M_A)=\delta_{J_A, M_A}$.
The off--shell
prescription used in all of the calculations has been the so--called
$CC1^{(0)}$ form.  This is the most used prescription and moreover,
the results obtained with it are not too different from the results obtained
with other prescriptions such as $CC2^{(0)}, NCC1$ and $NCC2$
for values of the
momentum $p$ not too high (for a detailed study of this subject see Ref.~[3]).
In the $CC1^{(0)}$ prescription
current conservation has been imposed by
eliminating the longitudinal components of the current in favor of the
charge components and the specific form of the single--nucleon current
is
$$\Gamma^{\mu}_{CC1}=(F_1+F_2)\gamma^{\mu}-{F_2\over 2M_N}
({\bar P}+P_N)^{\mu}\ \ , \eqno(45)$$
with ${\bar P}\equiv ({\bar E},{\bf p})$ the four--momentum
for kinematics having the same three--momentum as the struck
nucleon (${\bf p}$), but on--shell energy ${\bar E}=\sqrt{{\bf p}^2+M_N^2}$.
$F_1$ and $F_2$ are the on--shell Pauli and Dirac form factors, respectively.
They are related to the Sachs form factors in the
usual way:
$$\eqalign{
F_1(\tau) &=\left[G_M(\tau)-G_E(\tau)\right]/(1+\tau) \cr
F_2(\tau) &=\left[G_E(\tau)+\tau G_M(\tau)\right]/(1+\tau)\ \ , \cr}
\eqno(46)$$
where $\tau\equiv -Q^2/4M_N^2$.
Simple expressions have been assumed for the single--nucleon form factors,
{\it viz.}, dipole $\tau$--dependences for $G_{Mp}$, $G_{Mn}$ and $G_{Ep}$
together with the Galster$^{[32]}$
parametrization for $G_{En}$ (see Refs.~[33] for explicit parametrizations).

Before entering into a discussion of the different figures to follow,
we briefly summarize the kinematics involved in the process.
As noted in previous sections, our aim is to explore the different polarization
observables in the quasi--free region where one expects to be probing
essentially the single--nucleon content of the nucleus. Furthermore,
at sufficiently
high momentum transfer the process is expected to be only mildly influenced
by final--state interactions and various exchange effects not considered
in this work. Therefore, the kinematics are selected first by fixing the value
of the momentum transfer $q$ to be reasonably large. Two cases have been
considered: $q=500$ MeV/c and $q=1$ GeV/c. Then for both $q$--values the
energy transfer $\omega$ considered is set to the value corresponding to the
quasielastic peak
$$\omega_{QP}\equiv\{\sqrt{q^2+M_N^2}-M_N\}+E_s\ \ , \eqno(47)$$
where $E_s$ is the separation energy given by
$E_s\equiv M_N+M_B^0-M_A$.

Solving the energy--balance equation (Eq.~(2)) subject to the condition
$-1\leq \cos\theta \leq +1$, one can obtain the range of allowed values
of the bound nucleon momentum: $p_{min}\leq p\leq p_{max}$.  For instance,
from the developments in Ref.~[34] it can be shown that in the so--called
$y$--scaling region ($y\leq 0\leftrightarrow \omega\leq\omega_{QP}$) the
values $p_{min}$ and $p_{max}$ are given as follows:
$$\eqalignno{p_{min} &={1 \over{4 W^2}}\left\{ q(\Lambda_+ +\Lambda_-)-
2\sqrt{q^2+W^2}\sqrt{\Lambda_+ \Lambda_-} \right\} &(48a) \cr
p_{max} &={1 \over{4 W^2}}\left\{q (\Lambda_+ +\Lambda_-)+
2\sqrt{q^2+W^2}\sqrt{\Lambda_+\Lambda_-} \right\}\ \ , &(48b)\cr}$$
where we have introduced the quantities
$$\eqalignno{W^2&=(M_A+\omega)^2-q^2 &(49a)\cr
\Lambda_{\pm} &=(W\pm M_B^0)^2-M_N^2\ \ , &(49b)\cr}$$
so then $\Lambda_+ -\Lambda_- = 4 W M_B^0$; here $W$ is the total CM
energy.  The momentum of the ejected nucleon $p_N$, and the angle
$\theta$ defining the direction of ${\bf p}$ are given by
$$p_N=\sqrt{\left\{\omega+M_A-\sqrt{p^2+M_B^2}\right\}^2-M_N^2}
\eqno(50)$$ $$\cos\theta={p_N^2-q^2-p^2 \over 2pq}\ \ . \eqno(51)$$
Accordingly, for fixed values of $q$, $\omega$ and the mass of the
daughter nucleus ($M_B$) one can change the bound nucleon momentum $p$
in the interval between $p_{min}$ and $p_{max}$ by varying the angle
$\theta_N$.  Note that as $p$ varies, the final outgoing nucleon
momentum $p_N$ also varies and so does the angle $\theta$.  However, the
variation of $p_N$ is tiny for the whole range of variation of $p$ in
the case in which $M_B^2$ is large compared with $p^2$, as is the
general case for complex nuclei.  Accordingly, we have that $p_N$ is
slightly smaller than $q$ under typical conditions for the kinematics
studied here. On the contrary, the value of $\theta_N$ given by
$$\cos\theta_N={p_N^2+q^2-p^2 \over 2p_Nq} \eqno(52)$$ varies quite
rapidly as $p$ changes.

As a specific example, let us consider
the case in which the residual nucleus is in its
ground state. Choosing
$\omega=\omega_{QP}$ we get
$$\eqalignno{p_{min} & = 0 &(53a)\cr
p_{max}&= 2q \left[ {{ 1+\sqrt{q^2+M_N^2}/M_B^0 }
\over{ [1+\sqrt{q^2+M_N^2}/M_B^0]^2-[q/M_B^0]^2 }} \right]
\approx 2q\ \ , &(53b)\cr}$$
where the approximate result in Eq.~(53b) holds for $q<\!\!< M_B^0$.
Thus, $p$ can take all the possible values between 0 and $2q$. In this range
of variation of $p$, the values of $p_N$ vary between,
$$\eqalignno{p_N(p_{min})&=q &(54a)\cr
p_N(p_{max}) &=
q\left[1-{\cal O}\left({q\over M_B^0}\right)\right]\ \ . &(54b)\cr}$$
For the particular cases considered here, $q\leq 1$ GeV/c and
$M_B^0\approx 19$ GeV/c, the difference between the maximum and
the minimum values
of $p_N$ amounts to less than 5\% and hence, all of the effects
related to final--state interactions that depend on $p_N$ should
remain essentially constant.

In what follows we will present and discuss the results obtained (cross
sections, response functions and asymmetries) from various points of
view. First, we study the dependence of the observables on the
struck--neutron momentum $p$. Different orientations of the target
polarization are considered and a comparison between the results
obtained using Nilsson and Hartree--Fock single--particle wave functions
is presented. Second, we study the general properties of the observables
and their dependence on the target polarization orientation. In the two
cases, we discuss the situations in which the final nucleus is in its
ground state and also consider the effects of having contributions from
excited states.  Third, we also study the influence of the nuclear
deformation on the cross sections and asymmetries. Finally, we explore
the effects when one considers high values of the momentum transfer $q$;
specifically, we compare the results obtained for $q=1$ GeV/c with those
for $q=500$ MeV/c.

\goodbreak\bigskip
\noindent{\bf 5.3\quad The $p$--dependence of the Observables}
\medskip\nobreak
In this section we study the dependence of differential cross sections
and polarization observables on the momentum of the bound neutron $p$.
The results are presented in Figs.~9--16. The value of the momentum
transfer has been fixed to $q=500$ MeV/c and three different target
polarization directions have been considered (along $x$--, $y$-- and
$z$--axes) (see Fig.~2). Figures~9--11 correspond to the case in which
the residual nucleus $^{20}$Ne is in its ground state. As mentioned in
Sect.~4, for this case only the $\ell_j=d_{3/2}$ component of the
struck--neutron wave function enters. In Figs.~12--16 a sum over all the
single--particle quantum numbers allowed by transitions to the ground
and first excited state in $^{20}$Ne has been performed; the inclusion
of higher excited states in the daughter nucleus does not basically
change the results shown here.  In all of these figures the momentum $p$
covers the interval $0\leq p\leq 2$ fm$^{-1}$, while for higher values
of $p$ the cross section would be too small to be measured and
additionally short--range correlations (not included in our analysis)
may start to play an important role.  For the Nilsson model, the
oscillator parameter $b$ has been fixed to reproduce the DDHF radius of
the deformed single--nucleon orbital involved in the calculations: the
value obtained is $b=1.79$ fm, which is a little bit lower than the
value $b=1.82$ fm required to fit the r.m.s. charge radius of the whole
nucleus (see Sect.~5.1). As discussed in Ref.~[14] the use of a larger
value, $b=1.82$ fm, in the Nilsson model would produce in the cross
sections and response functions higher peaks together with simultaneous
displacements to the left (to lower values of the momentum $p$).

We begin the discussion by considering the totally unpolarized cross
sections (incident electron and target not polarized) for the final
nucleus in its ground state (Fig.~9).  Forward--angle ($\theta_e=30^o$)
and backward--angle ($\theta_e=150^o$) electron scattering is
considered.  In the former case, the main contributions come from the
longitudinal responses, whereas the contributions of the transverse
responses are dominant in the latter. The cross section obtained for
forward--angle electron scattering is approximately a factor 10 higher
than the cross section obtained in the backward--angle situation, a
result that is obviously connected with the very different Mott cross
sections obtained in the two situations.  As mentioned above, co--planar
($\phi_N=0^o$) and out--of--plane ($\phi_N=90^o$) kinematics have both
been considered and in the latter the longitudinal--transverse
interference response is zero.  For forward--angle electron scattering
there are appreciable differences between the differential cross
sections for $\phi_N=0^o$ and $\phi_N=90^o$ (the latter being of the
order of $\sim 1.3$ higher in the maximum), while for backward--angle
electron scattering the $\phi_N=0^o$ and $\phi_N=90^o$ results are much
closer.  This can be explained by taking into account the fact that the
relative contributions of the responses which depend on the angle
$\phi_N$ (responses $TL, TT$), compared with the pure longitudinal and
transverse responses (independent of $\phi_N$), are different for
forward-- and backward--angle electron scattering. In the first case
($\theta_e=30^o$), such contributions are important, whereas for
backward--angle scattering the purely transverse response function
strongly dominates and the influence of the interference terms is
small. Note that in the case of an ejected neutron the longitudinal
response is much smaller than the transverse response at the momentum
transfer considered (see Fig.~16).

Also seen in Fig.~9 are the results given by the two different nuclear
models --- Nilsson and Hartree--Fock. The difference between them is
significant, being $\sim$30\% higher for DDHF wave functions in the
vicinity of the maximum, due mainly to the different weight, although
also partly to the different shape of the $d_{3/2}$ component of the
struck--neutron wave function in the two models (see Table~3). In
addition, note that the peak in the cross section obtained with Nilsson
wave functions occurs for a $p$--value that is slightly higher than the
corresponding one in the DDHF case.  A slightly larger value of the
Nilsson model oscillator parameter would improve the agreement with the
DDHF result.

In Fig.~10 we show the results for the target
polarization ratio ${\cal P}_{\Sigma}$.
As discussed in Sect.~4, for the particular case where
the daughter nucleus is in its ground state, ${\cal P}_{\Sigma}$ is
independent of the nuclear model used in the calculation.
The general expression for ${\cal P}_{\Sigma}$
(Eq.~(39)) in the case of a
100\% polarized target with angular momentum $J_A=3/2$ and
parity $\pi=+1$ is simply reduced to ${\cal P}_{\Sigma}=-P_2(\cos\xi)$.
Two orientations of the
target polarization (along the $z$-- and $x$--axes) are shown.
The case of target polarization along the $y$--axis can be related to that
along the $x$--axis: in particular, for co--planar kinematics and
target polarization along the $y$--axis,
the value obtained for ${\cal P}_{\Sigma}$ is the same
as with target polarization along the $x$--axis and out--of--plane
($\phi_N=90^o$)
kinematics. The situation where the target polarization is
along the $y$--axis and $\phi_N=90^o$ is also equivalent to that with
target polarization along the $x$--axis and $\phi_N=0^o$.
In the case of the target polarization along the $z$--axis,
${\cal P}_{\Sigma}$ is independent of the azimuthal angle $\phi_N$.
The full line represents the case $\phi_N=0^o$ and target polarization
along the $z$--axis, while the dashed (dotted) line corresponds to $\phi_N=0^o$
($\phi_N=90^o$) and target polarization along the $x$--axis.
In the situation given by $\phi_N=90^o$ and ${\bf P}^*$ parallel to the
$x$--axis, ${\cal P}_{\Sigma}$ is a constant (${\cal P}_{\Sigma}=1/2$) because
$\cos\xi=0$ (see Eq.~(27)).
This is always the case when the target polarization vector
and the momentum of the ejected nucleon are in perpendicular planes and
${\bf P}^*$ is perpendicular to ${\bf q}$.
The response functions
$R^K$, $K=L,T,TL,TT$ obtained with 100\% polarized
$^{21}{\overrightarrow{\hbox{Ne}}}$,
are a factor 1.5 bigger than the responses obtained if the target was not
polarized. In the case where $\phi_N=0^o$ and ${\bf P}^*$ is parallel to the
$x$--axis, the ratio ${\cal P}_{\Sigma}$ presents a very different behaviour;
then one has $\cos\xi=\sin\theta$ and hence ${\cal P}_{\Sigma}$ depends on the
value of the momentum $p$ (see Eq.~(27)).
For the extreme value $p=0$ one has $\cos\xi=\pm 1$
and consequently ${\cal P}_{\Sigma}=-1$. As the value of $p$ increases,
${\cal P}_{\Sigma}$ decreases slowly in absolute value (its variation is of
the order of $\sim$20\% for the interval of momenta considered).
At $p\approx 0.75$ fm$^{-1}$, where the cross section is maximum,
${\cal P}_{\Sigma}\approx -0.95$. Therefore,
the ratio between the
electron--unpolarized cross sections $\Sigma$ for polarized and
unpolarized targets grows from zero to 0.2 in the interval from
$p=0$ to $p=2$ fm$^{-1}$.
Finally, in the case where ${\bf P}^*$ is parallel to the $z$--axis,
${\cal P}_{\Sigma}$ does
not depend on $\Delta\phi$ and one has ${\cal P}_{\Sigma}=1/2$ at $p=0$,
which decreases slowly with $p$, being of the order of 0.25 for
$p=2$ fm$^{-1}$.

In Fig.~11 the results for the electron--target polarization
ratio ${\cal P}_{\Delta}$ are shown. Figure~11(a) corresponds to
forward--angle electron scattering ($\theta_e=30^o$) and Fig.~11(b) to
backward--angle electron scattering ($\theta_e=150^o$). The orientations
of the target polarization vector and values of the angle $\phi_N$ have
been selected as discussed above. We start with
the results shown in Fig.~11(a). When
the target polarization is along the $z$--axis, the asymmetry
${\cal P}_{\Delta}$ depends very slightly on the value of $\phi_N$ selected.
The difference between the results obtained for $\phi_N=0^o$ and $\phi_N=90^o$
is at most of the order of $\sim$15--20\%.
The behaviour of ${\cal P}_{\Delta}$ is
the same in both cases, decreasing (towards zero) as the value
of the momentum $p$ increases. For the other two orientations of the
target polarization (along $x$-- and $y$--axes) the behaviour presented by
${\cal P}_{\Delta}$ depends much more on the value of $\phi_N$ selected.
For example, when the target polarization vector is oriented along the
$x$--axis and $\phi_N=0^o$ (co--planar kinematics), ${\cal P}_{\Delta}$
is very close to zero for small values of $p$ and increases as one goes to
higher values of $p$. The same type of behaviour, especially
for backward--angle electron scattering, is obtained when the target
polarization points along the $y$--axis and $\phi_N=90^o$.
Finally, for the target polarization
oriented along the $x$--axis and $\phi_N=90^o$, ${\cal P}_{\Delta}$ is
different from zero and almost independent of the value of the momentum $p$.
We shall see later that this last result is obtained even when one is not
restricted to the case of the residual nucleus in its ground state. The
situation with the target polarization along the $y$--axis ($N$--polarized
target) and co--planar kinematics gives ${\cal P}_{\Delta}=0$, as already
mentioned in Sect.~4.
The discussion of the results for backward--angle electron scattering
(Fig.~11(b)) follows the same trend. The only difference is the value
of ${\cal P}_{\Delta}$ obtained for some particular situations: for example,
when the target polarization is oriented along the $z$--axis,
${\cal P}_{\Delta}$ is approximately twice as small as the value obtained
in the previous case, whereas if $\theta^*=\phi^*=\phi_N=90^o$,
${\cal P}_{\Delta}$ is twice as large.

Figures~12-16 show the results obtained when the contributions from
the first excited state $J_B^{\pi}=2^+$ in the residual nucleus $^{20}$Ne
are also included.
The allowed quantum numbers of the struck--neutron wave function
in the spectral function are
$j,j'=3/2,5/2,7/2$ and $\ell, \ell'=2,4$. The odd--neutron wave function
contains all of these components and they carry practically all of the
normalization strength (see Table~3).
Figure~12 corresponds to the totally unpolarized cross section. Again,
forward-- and backward--angle electron scattering, as well as values of
$\phi_N=0^o$ and $\phi_N=90^o$, have been considered. The cross sections
obtained are approximately a factor 25--35 bigger than the results shown in
Fig.~9. It is important to remark that the main reason for this is that
the cross section for the transition to the $0^+$ state in $^{20}$Ne is
much smaller than that for the transition to the $2^+$ state,
due to the fact that in the former case only the $d_{3/2}$ component enters
carrying only 4--5\% of the normalization of the neutron wave function
(Table~3). The shapes are not so different because the dominant component
$d_{5/2}$ (which only enters in the latter case) has a similar shape to that
of the $d_{3/2}$ component. Similar arguments explain why
the differences produced by using Nilsson or
DDHF single--nucleon wave functions are much smaller in this case, being only
of the order of $\sim$2.5\% higher
for DDHF in the region close to the maximum.
The dependence on $\phi_N$ for forward-- and backward--angle electron
scattering is similar to the case shown in Fig.~9.

In Fig.~13 we show the results for the target polarization ratio
${\cal P}_{\Sigma}$. The situations selected are the same as discussed in
Fig.~10. The value of ${\cal P}_{\Sigma}$ for nuclei polarized along
the $x$--axis and $\phi_N=0^o (90^o)$ is the same as for $N$--polarized
nuclei and $\phi_N=90^o (0^o)$. The differences produced by using Nilsson
or DDHF wave functions are almost negligible for $p\leq 1$ fm$^{-1}$.
For higher values of $p$ such differences start to increase, being of the order
of $\sim$25\% when $p=1.5$ fm$^{-1}$. For $p > 1.5$ fm$^{-1}$, the results
in Fig.~13 must be viewed with caution because the tails of the
presently used single--particle
wave functions in momentum space are likely not reliable and,
as previously noted, in this
region dynamical short--range correlations begin to be important and to produce
strong enhancements of the high--momentum tails of the
Fourier transforms of the single--nucleon wave functions. Comparing the
results in Fig.~13 with the case where the residual nucleus is in its ground
state
(Fig.~10), one notes that ${\cal P}_{\Sigma}$ is now much smaller in absolute
value, {\it i.e.,\/} the effects introduced in the electron--unpolarized
response functions by the polarization of the target are much weaker.
In particular, at the cross section peak ($p\approx 0.75$ fm$^{-1}$)
these effects are of the order of
8\% for target polarization along ${\bf q}$ or target polarization
along the $x$--axis and $\phi_N=90^o$, and of the order of
15\% for target polarization along the $x$--axis and $\phi_N=0^o$.

The results for the asymmetry ${\cal P}_{\Delta}$ are
presented in Figs.~14 (forward--angle electron scattering) and
15 (backward--angle electron scattering).
Some of the comments previously made in the
discussion of Figs.~11 and ~13 can also be applied here and consequently we
only mention the most noteworthy differences. For the target polarized
along the $z$--axis, the dependence of ${\cal P}_{\Delta}$ in $\phi_N$ is
weak, especially for backward--angle electron scattering where there is
basically
no difference between the results obtained for $\phi_N=0^o$ and $\phi_N=90^o$.
Furthermore, ${\cal P}_{\Delta}$ does not
change much with momentum except for $p$ in the vicinity of
2 fm$^{-1}$ where
the Hartree--Fock wave functions have zeros. For
the other two selected orientations of the target polarization
(Figs.~14(b), ~14(c) and ~15(b), ~15(c)), one sees that
the behaviour of ${\cal P}_{\Delta}$ depends mainly on $\Delta\phi=\phi^*-
\phi_N$. For $|\Delta\phi|=90^o$, ${\cal P}_{\Delta}$ is almost constant
for all $p$--values and in particular for $N$--polarized nuclei
${\cal P}_{\Delta}=0$ as expected. For $\Delta\phi=0^o$, $|{\cal P}_{\Delta}|$
increases with the momentum $p$.

Finally, in Fig.~16 we show the results for the six hadronic response
functions $R^K, R^{K'}$ obtained with Nilsson and DDHF wave functions.
Co--planar kinematics and two orientations of the target polarization
(along the $z$-- and $x$--axes) have been selected. Note that the
results obtained with the two models are very similar for all of the
responses.  For high momenta where such differences start to increase
very quickly, the spectral function for transitions leaving the daughter
nucleus in discrete states fall off rapidly and therefore the response
functions and cross section are very small. For the
electron--unpolarized responses $R^K$, $K=L,T,TL,TT$, the difference
between the results obtained for the two orientations of the target
polarization is of the order of 25--30\% in the region close to the
maximum, $p\approx 0.75$ fm$^{-1}$. From the results in Fig.~13, one
sees that for higher values of the struck--neutron momentum $p$ such
difference starts to decrease, whereas for $p<0.8$ fm$^{-1}$ the
difference increases. The behaviour presented by the two
electron--polarized response functions is different: for $R^{T'}$ the
results for the two orientations of the target polarization are much
more different, of the order of a factor $\sim 6$ smaller (around the
peak) for the case where ${\bf P}^*$ is parallel to ${\bf q}$.  The fact
that a neutron is ejected in the scattering reaction explains why the
purely transverse response $R^T$ is dominant, {\it viz.,\/} a factor of
the order of $\sim 30$ bigger than the $R^L$--response in the region
close to the peak.

\goodbreak\bigskip
\noindent{\bf 5.4\quad Dependence of the Asymmetries
on the Target Polarization Orientation}
\medskip\nobreak
In Figs.~17--22 we show the results obtained
for the asymmetries ${\cal P}_{\Sigma}$ and
${\cal P}_{\Delta}$ as functions of the target polarization angle $\theta^*$.
The value of the momentum transfer has again been fixed to $q=500$ MeV/c and
the kinematics selected are the same as in the
previous figures, co--planar ($\phi_N=0^o$) and out--of--plane ($\phi_N=90^o$).
In all cases, DDHF single--nucleon wave functions have been used, and
five values of the bound neutron momentum $p$ have been
selected, $p=0.2, 0.6,1.0,1.4$ and $1.8$ fm$^{-1}$. The relationship between
these $p$--values and the angles $\theta$ and $\theta_N$ in the case where the
residual nucleus $^{20}$Ne is in its ground state can be seen in
Table~4. These $\theta$-- and $\theta_N$--values would only be slightly
different when contributions from excited states in $^{20}$Ne are included.
Also for simplicity in the discussion, in all cases we have
restricted the target polarization vector ${\bf P}^*$ to
the scattering plane, {\it i.e.,\/} $\phi^*=0^o$. Three possible
situations for
the residual nucleus are considered, i) ground state ($0^+$), ii) ground
state and first excited state ($0^+, 2^+$) and iii) ground state and first two
excited states ($0^+,2^+,4^+$).

We start the discussion with the case of ${\cal P}_{\Sigma}$. As seen in
Eq.~(38), the entire dependence of ${\cal P}_{\Sigma}$ on the target
polarization direction is given through the Legendre polynomials
$P_I(\cos\xi)$, where $\xi$ is the relative angle between the target
polarization vector ${\bf P}^*$ and the momentum of the bound neutron
${\bf p}$. $I$ is the tensor rank which can only take the value $I=2$ in
this case.  Therefore, the maximum value of $|{\cal P}_{\Sigma}|$ is
obtained when ${\bf p}$ and ${\bf P}^*$ are parallel or
antiparallel. Note that ${\cal P}_{\Sigma}$ positive means an increase
of the differential cross section due to target polarization, while
${\cal P}_{\Sigma}$ negative means a decrease, and that the sign of
${\cal P}_{\Sigma}$ does not only depend on $P_2(\cos\xi)$ but on the
specific transitions considered. Accordingly, for observational purposes
it is interesting to note which are the largest positive values that
${\cal P}_{\Sigma}$ takes on for different situations.  To analyze this
we consider the conditions for extrema in ${\cal P}_{\Sigma}$: for
arbitrary fixed values of $\Delta\phi$ these are $$\eqalignno{
\tan \theta^*= &\tan \theta \cos \Delta\phi &(55a)\cr
\cot \theta^*= &-\tan \theta \cos\Delta\phi\ \ .
&(55b)\cr}$$
For $\Delta\phi =0^o$ (${\bf P}^*$ and ${\bf p}_N$ located in the same
plane) condition (55a) corresponds to ${\bf p}$ parallel to ${\bf P}^*$
which is the condition for $|{\cal P}_{\Sigma}|$ maximum mentioned above;
on the other hand,
condition (55b) corresponds to ${\bf p}$ perpendicular to ${\bf P}^*$.
For $\Delta\phi=\pm 90^o$ (${\bf P}^*$ and ${\bf p}_N$ in perpendicular
planes) the extreme conditions are obtained for fixed values of $\theta^*$
independent of the direction in which the struck neutron is moving.
Condition (55a) corresponds to $\theta^*=0^o,180^o$ and
condition (55b) corresponds to $\theta^*=90^o$. This is ilustrated in
Figs.~17--19.

Figure~17 corresponds to the situation in which the residual nucleus is
in its ground state. In such a case one has ${\cal
P}_{\Sigma}=-P_2(\cos\xi)$. For $\phi^*=\phi_N=0^o$ ($\Delta\phi=0^o$)
(see Fig.~17(a)), the maximum in ${\cal P}_{\Sigma}$ is obtained for
$\theta^*=\theta-90^o$ (see Table~4) and its value (${\cal
P}_{\Sigma}=1/2$) does not depend on the momentum $p$.  The minimum
value of ${\cal P}_{\Sigma}$ (${\cal P}_{\Sigma}=-1$) is also
independent of $p$ and is obtained for $\theta^*=\theta$ which is the
condition for $|{\cal P}_{\Sigma}|$ maximum.  This last condition means
that when the bound neutron is moving along the direction of the target
polarization vector ${\bf P}^*$, the response functions $R^K$,
$K=L,T,TL,TT$ are zero.  In other words, in this case the differential
cross section for unpolarized electrons and polarized target is
zero. The relation, $\cos(\theta^*-\theta)=\pm 1/\sqrt{3}$ gives the
condition for which the asymmetry ${\cal P}_{\Sigma}$ is zero, {\it
i.e.,\/} the polarization of the target does not have any observable
effect on the responses $R^K$.  The results in Fig.~17(b) correspond to
$\phi^*=0^o$ and $\phi_N=90^o$ ($\Delta\phi=-90^o$). As one can see, the
maximum in ${\cal P}_{\Sigma}$ is obtained for $\theta^*=90^o$ and its
value (${\cal P}_{\Sigma}=1/2$) is constant for all the struck--neutron
momenta $p$. The minimum value of ${\cal P}_{\Sigma}$ corresponds to
$\theta^*=0^o,180^o$ and its value depends on $p$ (see Eq.~(39)) through
the Legendre polynomial.  Finally, the zeros of ${\cal P}_{\Sigma}$ are
given by the relation, $\cos\theta^*\cos\theta=\pm 1/\sqrt{3}$, which
can only be satisfied at $p$--values larger than those considered in
Fig.~17(b).

Figures~18--19 show the results for ${\cal P}_{\Sigma}$ including the
contributions from the first and also from the first and second excited
states in the residual nucleus $^{20}$Ne, respectively. Contrary to the
previous situation, in these cases the extreme values of ${\cal
P}_{\Sigma}$ always depend on the value of the struck--neutron momentum
$p$.  The conditions for maximum and minimum depend not only on $p$ but
also on the number of excited states of the residual nucleus considered
in the calculations. In this sense, compare for example the results at
$p=1.4$ fm$^{-1}$ and 1.8 fm$^{-1}$ in Figs.~18(a) and 19(a) or 18(b)
and 19(b). As seen in Figs.~18(b) and 19(b), for $\phi^*=0^o$ and
$\phi_N=90^o$, ${\cal P}_{\Sigma}$ remains small and changes little with
$\theta^*$ for fixed $p$. As a general rule if one compares the results
in Figs.~18--19 with the previous case shown in Fig.~17, one can see
that the range of values of ${\cal P}_{\Sigma}$ is now at least a factor
10 smaller.  This effect was already discussed in Sect.~5.1 and allows
us to conclude that the poorer the energy resolution is ({\it i.e.,\/}
when more states of the residual nucleus are involved in the analysis),
the less the effects of the target polarization will be. The magnitude
of this effect also depends on the nuclear model used to describe the
nuclear excitation.

In Figs.~20--22 we present the results for the polarization ratio ${\cal
P}_{\Delta}$. Only results for forward--angle electron scattering
($\theta_e=30^o$) are shown in the figures, although the results for
backward--angle electron scattering are similar except for the fact that
the maximum values of ${\cal P}_{\Delta}$ are generally larger (by a
factor of the order of 2 for $\theta_e=150^o$), as illustrated
previously in Figs.~11, 14 and 15.  As seen in Sect.~4, the general
expression for ${\cal P}_{\Delta}$ (Eqs.~(40, 41)) is much more complex
than in the case of ${\cal P}_{\Sigma}$. The latter asymmetry only
depends on the relative angle between ${\bf p}$ and ${\bf P}^*$, while
${\cal P}_{\Delta}$ depends in addition on ${\bf p}_N$.  Moreover, for
$^{21}{\overrightarrow{\hbox{Ne}}}$ there are two tensor components that
contribute to ${\cal P}_{\Delta}$ with ranks $I=1$ and $I=3$, and
therefore, there are no simple analytical expressions for the extreme
conditions of ${\cal P}_{\Delta}$ or for their zeros.  However, when
$|{\cal P}_{\Delta}|$ is large the component $I=1$ strongly dominates
(in these cases the contribution of $I=3$ is at most of the order of
$\sim$20\%) and the extreme conditions obtained considering only the
component $I=1$ reproduce the behaviour seen in Figs.~20--22 reasonably
well.

Figure~20 corresponds to the case where the residual nucleus is in its
ground state.  For $\phi_N=0^o$ (Fig.~20(a)), ${\cal P}_{\Delta}$
reaches its maximum for small values of $\theta^*$. As the momentum $p$
increases the maximum of ${\cal P}_{\Delta}$ is slightly displaced to
higher values of $\theta^*$.  Note that for $0.2 \leq p\leq 1.8$
fm$^{-1}$ the maxima of ${\cal P}_{\Delta}$ are obtained for $\theta^* <
35^o$.  As $\theta^*$ increases, ${\cal P}_{\Delta}$ decreases and
becomes very close to zero for $\theta^*\approx 110^o$. In this last
situation, the polarization of the incident electrons has no observable
effect on the differential cross section. For $\phi_N=90^o$ (Fig.~20(b))
$|{\cal P}_{\Delta}|$ reaches its maximum for $\theta^*=0^o$ and
$\theta^*\approx 180^o$, independent of the value of the momentum $p$.
As the $p$--value increases, the maximum value of $|{\cal P}_{\Delta}|$
decreases. It is interesting to note that for $\theta^*=90^o$ the values
of ${\cal P}_{\Delta}$ are almost the same ($\approx -0.1$) for all
momenta $p$ selected.

In Figs.~21--22 contributions from excited states in the residual
nucleus $^{20}$Ne have been added to the contributions to the
ground state ($0^+$), in particular, the first ($2^+$) excited state in
Fig.~21, and the first and second ($4^+$) excited states in Fig.~22. In both
cases one sees that ${\cal P}_{\Delta}$ has a similar dependence on
$\theta^*$ and also
takes on similar values. For $\phi_N=0^o$ (Figs.~21(a)--22(a)) one can
clearly observe different regions: for $\theta^* \leq 60^o$,
$|{\cal P}_{\Delta}|$ is rather constant and significant;
for $60^o \leq \theta^* \leq 130^o$, $|{\cal P}_{\Delta}|$
varies quite rapidly with $\theta^*$ until it reaches its maximum
in the range $130^o \leq \theta^* \leq 150^o$.
For $\phi_N=90^o$ (Figs.~21(b)--22(b)), ${\cal P}_{\Delta}$ has a very
regular behaviour that is independent of the $p$--value: for
$\theta^*\approx 90^o$,
${\cal P}_{\Delta}$ is approximately zero and for $\theta^*\approx 0^o, 180^o$,
$|{\cal P}_{\Delta}|$ takes on its maximum value.
In summary, in the different kinematical situations considered, the absolute
values of the asymmetry ${\cal P}_{\Delta}$ are always more important for
$\theta^*$--values close to $0^o$ and $180^o$.

To finish with the presentation of the results in this section,
in Fig.~23 we show the total differential
cross section (Eq.~(1)) as a function of
the target polarization angle $\theta^*$.
Forward--angle electron scattering ($\theta^*=30^o$)
has been selected and contributions from the ground and first excited state
in the residual nucleus have been included in the calculations.
The values of the momentum transfer $q$ and the
struck--neutron momentum $p$ have been fixed to $q=500$ MeV/c and
$p=0.75$ fm$^{-1}$, respectively, where, as previously stated,
the latter corresponds approximately
to the momentum where the peak of the cross section occurs. For the kinematics
considered, $\theta=100.14^o$.
Three values of the azimuthal angle $\phi_N$ have been used,
$\phi_N=0^o$ (Fig.~23(a)), $\phi_N=66^o$ (Fig.~23(b)) and $\phi_N=90^o$
(Fig.~23(c)).
In each case, the differential cross sections obtained
for different orientations of the target polarization are represented and
the totally
unpolarized cross section is also shown for comparison. Note that the
general behaviour of the cross section is similar in the three graphs.
{}From the results obtained, one can see that
the maximum in the total cross section is always reached when $\phi^*=\phi_N$,
{\it i.e.,\/} when the target polarization vector ${\bf P}^*$
and the ejected neutron momentum ${\bf p}_N$
are located in the same plane. Any other choice for the angles
$\phi^*$ and $\phi_N$ reduces the maximum of the cross section. These results
also allow one to see
which are the target polarization and out--going--nucleon directions that
produce a higher increment of the differential cross section relative to
the unpolarized case. Specifically, for the selected kinematics (in Figs.~23,
$\theta_N\approx 15^o$ and $\theta\approx 110^o$) the highest increment
corresponds to $\phi_N=\phi^*$ and $\theta^*\approx 130^o$.

\goodbreak\bigskip
\noindent{\bf 5.5\quad Dependence on the Nuclear Deformation}
\medskip\nobreak
In this section we briefly discuss the effect that the nuclear deformation
has on
the various observables above discussed.
For this purpose we use the Nilsson model where
we can easily see what is the effect of deformation by changing the value
of $\delta$. The results are presented in Fig.~24:
for both unpolarized cross sections (Fig.~24(a)) and asymmetries
${\cal P}_{\Sigma}$ and ${\cal P}_{\Delta}$ (Fig.~24(b)), we present
ratios between the results obtained for different values of the deformation
parameter $\delta$ and the result corresponding to the equilibrium deformation
$\delta= 0.31$ (see Sect.~4).
All of the results correspond to forward--angle electron
scattering ($\theta_e=30^o$), co--planar kinematics ($\phi_N=0^o$) and
target polarization ${\bf P}^*$ oriented along ${\bf q}$--direction.
Contributions from the ground and first excited state in the residual
nucleus $^{20}$Ne have been included in the calculations. It should be
remarked that in the case of the residual nucleus in its ground state
the asymmetries ${\cal P}_{\Sigma}$ and ${\cal P}_{\Delta}$ are not affected
by the nuclear deformation. Figure~24(a) shows the variation of the unpolarized
cross
section with the nuclear deformation for three different values of the bound
neutron momentum, $p=0.5$ fm$^{-1}$ (full), 0.75 fm$^{-1}$ (short--dashed)
and 1.5 fm$^{-1}$ (dashed). As observed in the figure, the weakest
dependence on the nuclear deformation is obtained
for $p=0.75$ fm$^{-1}$ which corresponds to the maximum in
the cross section. On the contrary, for large $p$--values the cross section
falls as low as a factor of four below
the equilibrium result for very large deformations
($\delta=0.7$).
Figure~24(b) corresponds to the asymmetries ${\cal P}_{\Sigma}$ and
${\cal P}_{\Delta}$.
One can observe that the effects of the nuclear deformation lie within
a range of $\sim$25--30\% in three cases. Only for $p=1.5$ fm$^{-1}$
and ${\cal P}_{\Sigma}$ do the results clearly deviate from the equilibrium
value --- in this case because of a change in the sign of ${\cal P}_{\Sigma}$
when one goes to large values of the deformation parameter $\delta$.
\goodbreak\bigskip
\noindent{\bf 5.6\quad Results for High Momentum Transfer}
\medskip\nobreak
All of the results presented in previous figures have been obtained by fixing
the value of the momentum transfer to $q=500$ MeV/c and fixing the energy
transfer $\omega$ to the value given by Eq.~(47) corresponding to the
quasielastic peak.
In this section our aim is to study briefly the possible effects
introduced when one considers a larger $q$--value, still working in the
region of the quasielastic peak. In particular, we fix $q=1$ GeV/c where one
expects that for such high momentum transfer the assumptions that go into
the PWIA will become even more valid. Specifically, in Figs.~25 we show
the ratios between the results obtained for
$q=1$ GeV/c and $q=500$ MeV/c: Fig.~25(a) corresponds to
the totally unpolarized cross section and Figs.~25(b) and 25(c) to
the asymmetries ${\cal P}_{\Sigma}$ and ${\cal P}_{\Delta}$, respectively.
Contributions from the ground
and first excited state in the residual nucleus $^{20}$Ne have been included in
the calculations. As usual, co--planar ($\phi_N=0^o$) and out--of--plane
($\phi_N=90^o$) kinematics, as well as three orientations --- along the
$x$--, $y$-- and $z$--axes --- have been considered for the target
polarization.
Before entering into a discussion of the results, it is interesting to remark
that for a fixed value of the struck--neutron momentum $p$, the corresponding
value of the angle $\theta_N$ defining the direction of the ejected neutron
momentum is much smaller for higher values of $q$
than in the situation treated above (see Eq.~(52)) yielding,
for example, $\theta_N=47.06^o$ for $q=500$ MeV/c and $\theta_N=22.85^o$
for $q=1$ GeV/c at $p=2$ fm$^{-1}$.

We start the discussion with the results for the totally unpolarized cross
section (Fig.~25(a)). Full and dashed lines correspond to $\phi_N=0^o$ and
$\phi_N=90^o$, respectively. Note that in both cases the results vary only
mildly with the struck--neutron momentum $p$, with the variation being
at most on the order of
6--7\%. These results show that the totally unpolarized cross section
is roughly four times smaller for
$q=1$ GeV/c than for $q=500$ MeV/c at all $p$--values.
In Fig.~25(b) we present the results for the target polarization ratio
${\cal P}_{\Sigma}$. The three curves shown correspond
to $\phi_N=0^o$ and target polarization vector ${\bf P}^*$
along the $x$-- (full line), $y$-- (dashed) and $z$--axes (dotted).
As discussed in Sect.~5.1, the results for $\phi_N=90^o$ and the three
orientations selected of the target polarization would be similar.
Before entering into the discussion of the different specific situations, one
can see by examining the general expressions introduced in Sect.~4 (Eq.~(38))
that the difference in the asymmetry ${\cal P}_{\Sigma}$ obtained for
different values of the momentum transfer $q$ comes exclusively through the
Legendre polynomials $P_I(\cos\xi)$. For ${\bf P}^*$ parallel to the $y$--axis,
a constant ratio of unity is
obtained in the whole range of $p$--momenta, since
$\cos\xi=0$ and therefore ${\cal P}_{\Sigma}$ is independent of $q$. We
recall that in this situation
the response functions $R^{T'}$ and $R^{TL'}$ are zero and hence
${\cal P}_{\Delta}=0$.
For the other two orientations of the target polarization, we have
$\cos\xi=\cos\theta$ for ${\bf P}^*$ parallel to ${\bf q}$, and
$\cos\xi=\sin\theta$ for ${\bf P}^*$ parallel to the $x$--axis.
Accordingly, the ratio between the asymmetries ${\cal P}_{\Sigma}$ calculated
for $q=1$ GeV/c and $q=500$ MeV/c changes with the value of the struck--neutron
momentum $p$.
As $p$ increases, the ratio between the results obtained for
the two values of the momentum transfer $q$ also increases and one sees that
this effect is more
pronounced in the case when the target polarization vector is parallel to
${\bf q}$.

Finally, in Fig.~25(c) we present the results for the asymmetry
${\cal P}_{\Delta}$ where the scattering angle has been fixed to
$\theta_e=30^o$. For $\phi_N=0^o$ the results shown correspond to
${\bf P}^*$ parallel to ${\bf q}$ (dashed line)
and ${\bf P}^*$ parallel to the $x$--axis (full line), while for $\phi_N=90^o$
the curves correspond to the three orientations of the target polarization,
along the $x$-- (dotted), $y$-- (short--dashed) and $z$--axes (dot--dashed).
The behaviour presented by the results for $\phi_N=0^o$ and ${\bf P}^*$
parallel to the $x$--axis is due to the zero of the asymmetry
${\cal P}_{\Delta}$
for $p\approx 1.2$ fm$^{-1}$ and $q=500$ MeV/c (see Fig.~14(b)). For
$\phi_N=90^o$ and ${\bf P}^*$ parallel to the $y$--axis (short--dashed), the
results obtained are contained in the range $\sim$ [0.3--0.8]. In the region
$p<0.6$ fm$^{-1}$ the ratio decreases with the momentum $p$, whereas for
$0.6<p<1.0$ fm$^{-1}$ the ratio is rather constant ($\approx 0.3$) and
starts to increase slowly for higher $p$--values. In the case of the target
polarized along ${\bf q}$, the results obtained for $\phi_N=0^o$ and
$\phi_N=90^o$ are almost identical and equal to unity, indicating that in this
situation for a fixed value of the momentum $p$ the asymmetry
${\cal P}_{\Delta}$ takes on the same value regardless of the
value of the momentum transfer
$q$. Only for very high $p$--values, $p>1.8$ fm$^{-1}$, does the ratio start to
deviate slightly from unity. Finally, in the case where $\phi_N=90^o$ and
${\bf P}^*$ is parallel to the $x$--axis the
ratio (of the order of 1.2) is practically constant for all $p$--values. This
means that ${\cal P}_{\Delta}$ for $q=1$ GeV/c is of the order of
$\sim$20\% bigger than the value of ${\cal P}_{\Delta}$
obtained for $q=500$ MeV/c for a fixed value of $p$.
\goodbreak\bigskip
\noindent{\bf 6.\quad CONCLUSIONS}
\medskip\nobreak
In this work we have studied polarization degrees of freedom in coincidence
electron scattering from deformed nuclei within the context of the
Plane--Wave Impulse Approximation (PWIA). In particular, we have focused on
$(e,e'N)$ reactions where both the incident electrons and the target nuclei
are assumed to be polarized. Our main objective in this work has been centered
on a treatment of the polarized spectral function for deformed nuclei, while
a study of the
electron--nucleon cross section and single--nucleon response functions for
polarized electron scattering from bound ``off--shell" nucleons has already
been presented in Refs.~[3] and summarized in Sect.~2. As an initial choice
of deformed nucleus we have selected $^{21}$Ne for several reasons:
firstly, from an experimental point of view, $^{21}$Ne is a noble gas and,
as in the more familiar case of $^3\overrightarrow{\hbox{He}}$, is a
good candidate for a polarized target; secondly, it presents a
well--established
rotational energy spectrum;$^{[5]}$ and finally, its charge ($Z=10$) is not
large and therefore the electron Coulomb distortion will not be particularly
important in the description of the scattering. The PWIA is expected
to be a good approximation in the analysis of the reaction for the
sufficiently high momentum transfers and quasielastic kinematics
chosen throughout this work.

A general expression for the polarized spectral function for transitions
leaving the residual nucleus in discrete states
has been obtained. General properties and angular
symmetries have been discussed in detail in Sect.~3. Explicit expressions
for the hadronic response functions and polarization ratios have been
presented in Sect.~4 by combining the polarized spectral function with
the single--nucleon responses.$^{[3]}$ In this context, the asymmetry
${\cal P}_{\Sigma}$ (which depends exclusively on the polarization of the
target) has been proven not to be
affected by the ``off--shell" uncertainties in the treatment of the bound
nucleons. On the other hand, in the case of the residual nucleus in its
ground state, the two asymmetries introduced, ${\cal P}_{\Sigma}$ and
${\cal P}_{\Delta}$ have been proven to be independent of the nuclear model
used in the evaluation of the single--particle wave functions.

In obtaining specific results we have considered different representative
ranges of kinematics --- moderate and high $q$--values although always at the
quasielastic peak, forward-- and backward--angle
electron scattering, coplanar and non--coplanar nucleon detection and finally
different orientations of the target polarization axis ${\bf P}^*$. The
results presented in Sect.~5 for unpolarized
cross sections as well as for polarization observables (asymmetries
${\cal P}_{\Sigma}$, ${\cal P}_{\Delta}$) have been examined in detail from
different points of view. In general, we find that the range of predictions
obtained for such asymmetries is smaller than that seen for the individual
response functions or cross sections. We have shown the dependence of the
observables on the struck--neutron momentum $p$ and have seen that the
results obtained by
using Nilsson and DDHF single--nucleon wave functions are very similar
for $p$--values within the range [0--2] fm$^{-1}$. A general study of the
dependence of the polarization observables on the polarized target
orientation has also been performed and, in particular, we have explored
the kinematical situations for
which one observes the biggest effects due to the polarization of the target
and electrons. It should be noted that, as a general rule, one can conclude
that the poorer
the energy resolution is ({\it i.e.,\/} when more states of the residual
nucleus are involved in the analysis), the less will be the effects of
the target polarization.
Another interesting result obtained indicates that the global maximum in the
differential cross section is always reached when the target polarization
vector ${\bf P}^*$ and the ejected neutron momentum ${\bf p}_N$ are located
in the same plane. Furthermore, the dependence on the nuclear deformation
has been shown not to be very important (the effects are bigger for high
values
of the struck--neutron momentum $p$). Finally, we have shown that under some
specific kinematics, the polarization observables are almost independent of
the value considered for the momentum transfer $q$.

As already mentioned, in this work we have investigated in detail spin degrees
of freedom in coincidence electron scattering from deformed
nuclei within the context of the PWIA.
In future work, we intend also to present results for other interesting
targets, where we will focus on a
study of the interferences between different single--nucleon orbitals
and their connections with the polarization degrees of freedom, and to
extend the results presented here to the high--missing--energy regime for
application in studies of polarized $^3$He.  Our longer--term objective
is also to go beyond the PWIA.  In this regard, we end by remarking that
ultimately, if polarization--coincidence measurements of the type
suggested by the present studies were to prove feasible and if sufficiently
fine information on the various components of the polarized spectral function
could be obtained, then it would be possible to investigate not only the
nuclear structure issues involved in arriving at those distributions, but
also the nature of the final--state propagation of the outgoing nucleon.
Indeed, using a single species of deformed nucleus might provide the ideal
situation for such studies: then the particular nuclear configurations
could be selected and yet, as the distribution of nuclear matter in this
case is asymmetric, different aspects of the final--state interaction
could be probed --- for instance, the degree of ``transparency'' in different
directions could be explored.

\goodbreak\bigskip
\centerline{\bf ACKNOWLEDGEMENT}\medskip
Two of us (J.A.C. and G.I.P.) wish to thank the members of CTP (MIT) for their
hospitality. This work has been supported in part by DGICYT (Spain) under
contract No. PB92/0021--002--01 and by funds provided by the U.S. Department
of Energy (D.O.E.) under contract \#DE-AC02-76ER03069. One of us (E.G.) wishes
to thank MEC for support.

\goodbreak\bigskip
\noindent{\bf APPENDIX A.\quad POLARIZED SPECTRAL FUNCTION}
\medskip\nobreak
In this appendix we present in detail the algebra needed to obtain
a general expression for the polarized spectral function
in bound nuclear systems. The general expression for the polarized
spectral function is given by Eq.~(7).  There
the single--nucleon creation operators $a^+_{{\bf p}m}$ can be expanded over a
basis of irreducible tensorial operators $a^+_{\ell jm_j}(p)$
in the following way:$^{[2]}$
$$a^+_{{\bf p}m}=\sum_{\ell j}\sum_{m_\ell m_j}<\ell m_\ell
{1\over 2}m|jm_j>
Y_\ell^{*m_\ell}(\Omega)a^+_{\ell jm_j}(p)\ \ , \eqno(A.1)$$
where $<\ell m_\ell {1\over 2} m|jm_j>$
are the Clebsch--Gordan coefficients; $\ell,j$ are the single--nucleon
quantum numbers and $m_{\ell}, m_j$ their projections referred to
${\bf q}$--direction. $Y_{\ell}^{m_\ell}(\Omega)$ is
the spherical harmonic given by the angular variables $\Omega\equiv
\{\theta,\phi\}$
which define the direction of the momentum
${\bf p}$ in the $xyz$--system (Fig.~2).
In the case of annihilation operators $a_{{\bf p}m}$, the above expansion
reads
$$a_{{\bf p}m}=\sum_{\ell j}\sum_{m_\ell m_j}<\ell m_\ell
{1\over 2}m|jm_j>
Y_\ell^{m_\ell}(\Omega)(-1)^{j-m_j}
\tilde{a}_{\ell jm_j}(p)\ \ , \eqno(A.2)$$
where $\tilde{a}_{\ell j-m_j}$ are hole creation operators and consequently
irreducible
tensorial operators. They are related with
the annihilation operators $a_{\ell jm_j}$ by
$$\tilde{a}_{\ell jm_j}=(-1)^{j+m_j}a_{\ell j-m_j}\eqno(A.3)$$
Introducing these relations into the general expression for the spectral
function (Eq.~(7)) we obtain
$$\eqalign{&S_{mm'}({\bf p},E_m,\Omega^*)
=\sum_B\sum_{\ell\ell'}\sum_{jj'}
\sum_{m_\ell m'_\ell}\sum_{m_j m'_j} Y_{\ell}^{m_\ell}(\Omega)
Y_{\ell'}^{* m'_\ell}(\Omega)\cr
&\times <\ell m_\ell {1\over 2}m|jm_j>
<\ell'm'_\ell{1\over 2}m'|j'm'_j>^{*}
\sum_{M_AM_B}p(A)\sum_{M'_AM''_A}
{\cal D}_{M_AM'_A}^{* J_A}(\Omega^*)
{\cal D}_{M_AM''_A}^{J_A}(\Omega^*) \cr
&\times <J_AM'_A|a^+_{\ell jm_j}(p)|J_BM_B>^*
<J_AM''_A|a^+_{\ell'j'm'_j}(p)|J_BM_B>
\delta(E_m+\epsilon_B-\epsilon_A^0)\ \ . \cr}\eqno(A.4)$$
Expressing the Clebsch--Gordan coefficients in terms of the 3--j's and using
the Wigner--Eckart theorem and some relations held by the rotation matrices
(Ref.~[25]) we can write
$$\eqalign{
&S_{mm'}({\bf p},E_m,\Omega^*)
=\sum_B \sum_{\ell\ell'}\sum_{jj'}
\sum_{m_\ell m'_\ell}\sum_{m_j m'_j}
Y_{\ell}^{m_\ell}(\Omega)
Y_{\ell'}^{* m'_\ell}(\Omega)
(-1)^{\ell-\ell'+m_j-m'_j}[j][j'] \cr
&\times \sum_{IM}
\sum_{M_A}
p(A)(-1)^{J_A-M_A}[I]^2
{\cal D}^{*I}_{0 M}(\Omega^*)C^{*J_AJ_B}_{\ell j}(p)C^{J_AJ_B}_{\ell 'j'}(p)\cr
&\times \left( \matrix{ \ell & 1/2 & j \cr m_\ell & m & - m_j
\cr}\right) \left( \matrix{ \ell' & 1/2 & j'\cr m'_\ell & m' & -m'_j\cr}
\right) \left( \matrix{ J_A & J_A & I \cr M_A & -M_A & 0 \cr} \right) \cr
& \times \sum_{M'_AM''_AM_B}
(-1)^{J_A-M''_A}\left( \matrix{ J_A& J_A & I \cr M''_A & - M'_A & M
\cr}\right)
\left(\matrix{ J_A & j & J_B \cr -M'_A & m_j & M_B\cr}\right) \left(
\matrix{ J_A & j' & J_B\cr-M''_A & m'_j & M_B \cr}\right)\cr
&\times \delta(E_m+\epsilon_B-\epsilon_A^0)\ \ , \cr}\eqno(A.5)$$
where we have introduced the reduced nuclear matrix elements
$C^{J_AJ_B}_{\ell j}(p)$ and $C^{J_AJ_B}_{\ell'j'}(p)$ with
$C^{J_AJ_B}_{\ell j}(p)\equiv <J_A||a^+_{\ell j}(p)||J_B>$
and are using the notation
$[j]\equiv\sqrt{2j+1}$.

Summing over the indices $M'_A$, $M''_A$, $M_B$ and
using the relation (6.2.8) in Edmonds,$^{[25]}$ the expression of the spectral
function is reduced to
$$\eqalign{
&S_{mm'}({\bf p},E_m,\Omega^*)
=\sum_B\sum_{\ell\ell'}\sum_{jj'}
\sum_{m_\ell m'_\ell}\sum_{m_j m'_j}
\sum_{IM}f^{J_A}_{I}
(-1)^{\ell-\ell'-m'_j+J_A+J_B+M}[j][j'][I]
\cr
&\times {\cal D}_{0 -M}^{I}(\Omega^*)Y_{\ell}^{m_\ell}(\Omega)
Y_{\ell'}^{* m'_\ell}(\Omega)
C^*_{\ell j}(p)C_{\ell 'j'}(p)\cr
& \times \left(\matrix{\ell&1/2&j\cr m_\ell&m&-m_j\cr}\right) \left(
\matrix{ \ell'&1/2&j'\cr m'_\ell&m'&-m'_j\cr}\right)
\left( \matrix{ j'&j&I\cr -m'_j&m_j&-M\cr}\right)
\left\{\matrix{j'&j&I\cr J_A&J_A&J_B\cr}\right\}\cr
&\times
\delta(E_m+\epsilon_B-\epsilon_A^0)\cr}\eqno(A.6)$$
with $f_I^{J_A}$ being the spherical Fano statistical tensor given by
$$f^{J_A}_{I}= \sum_{M_A}p(A)(-1)^{J_A-M_A}[I]\left( \matrix{
J_A&J_A&I\cr M_A&-M_A&0\cr}\right)\ \ . \eqno(A.7)$$
Using the relations obeyed by the spherical harmonics and expressing the
rotation matrix containing the target polarization angles in terms of spherical
harmonics$^{[25]}$ one obtains
$$\eqalign{
&S_{mm'}({\bf p},E_m,\Omega^*)
=\sum_B\sum_{\ell\ell'}\sum_{jj'}\sum_{IM} f^{J_A}_{I}\sum_{LH}
(-1)^{\ell-\ell'+J_A+J_B}[j][j']
[\ell][\ell'][L]C^{*J_AJ_B}_{\ell j}(p)C^{J_AJ_B}_{\ell'j'}(p)\cr
&\times Y_{I}^{-M}(\Omega^*)Y_{L}^{-H}(\Omega)
\left(\matrix{ \ell&\ell'&L\cr 0 & 0 & 0\cr}\right)
\left\{\matrix{j'&j&I\cr J_A&J_A&J_B\cr}\right\}
(-1)^{-m}
\sum_{m_\ell m'_\ell}\sum_{m_j m'_j}
\cr
&\times\left( \matrix{\ell&1/2&j\cr m_\ell&m&-m_j\cr}\right)
\left(\matrix{ \ell'&1/2&j'\cr m'_\ell & m' &-m'_j\cr}\right)
\left(\matrix{ j'&j&I\cr -m'_j&m_j&-M\cr}\right)
\left(\matrix{ \ell&\ell'&L\cr m_\ell&-m'_\ell&H\cr}\right)\cr
& \times\delta(E_m+\epsilon_B-\epsilon_A^0)\ \ . \cr}\eqno(A.8)$$
Finally, one can even simplify the sum over the indices $m_\ell,\,m'_\ell,\,
m_j$ and $m'_j$ by using the relation (2.3.1) in Rotenberg$^{[35]}$. Taking
into account selection rules embodied in the 3--j and 6--j coefficients,
the final expression obtained for the different tensor components of the
polarized spectral function (Eq.~(16)) is
$$\eqalign{&S_{mm'}^{(I)}({\bf p},E_m,\Omega^*)
=(-1)^{m-1/2}f_I^{J_A}
\sum_{\ell\ell'}\sum_{jj'} \sum_{LK} \sum_{M}
(-1)^{J_A+J_B+j+\ell'}
[j][j'][\ell][\ell'][L][K]^2\cr
&\times C^{*J_AJ_B}_{\ell j}(p)C^{J_AJ_B}_{\ell'j'}(p)
Y_{I}^{-M}(\Omega^*)Y_{L}^{-H}(\Omega)
\left( \matrix{\ell&\ell'&L\cr0&0&0\cr}\right)
\left(\matrix{ L&I&K\cr H&M&N\cr}\right)\cr
&\times
\left(\matrix{K&1/2&1/2\cr N&m&-m'\cr}\right)
\left\{\matrix{J_A&J_A&I\cr j&j'&J_B\cr}\right\}
\left\{ \matrix{L&I&K\cr\ell&j&1/2\cr\ell'&j'&1/2\cr}\right\}
\delta(E_m+\epsilon_B-\epsilon_A^0)\ \ . \cr}\eqno(A.9)$$
\goodbreak\bigskip
\noindent{\bf A.1.\quad The Unpolarized Case}
\medskip\nobreak
{}From the above expression for the
polarized spectral function
we can easily obtain the results corresponding
to unpolarized target and unpolarized electron.
In such a case, the only tensor polarization component which contributes
in the spectral function is
$I=0$. The Fano statistical tensor is
simply $f^{J_A}_0=1/[J_A]$ regardless of the detailed population of the
magnetic substates.
Therefore, the spin--components of
the spectral function are reduced to
$$\eqalign{S_{mm'}({\bf p},E_m,\Omega^*) &=
\sum_B\sum_{\ell\ell'}\sum_{jj'}\sum_L (-1)^{j+\ell-\ell'+m'}
{[\ell][\ell'][L]\over [J_A]^2\sqrt{4\pi}}
Y_L^{-H}(\Omega)C^{*J_AJ_B}_{\ell j}(p)C^{J_AJ_B}_{\ell' j}(p)\cr
&\times \left(\matrix{\ell&\ell'&L\cr0&0&0\cr}\right)
\left(\matrix{L&1/2&1/2\cr-H& m&-m' \cr}\right)
\left\{\matrix{1/2&1/2&L\cr\ell'&\ell&j\cr}\right\}
\delta(E_m+\epsilon_B-\epsilon_A^0)\ \ .\cr}\eqno(A.10)$$
{}From the selection rules embodied in the 3--j coefficients one sees that
the only possible values of $L$ are 0 or 1.
Moreover, from parity considerations (as mentioned above, we consider only
parity--conserving processes)
the values of $\ell$ and $\ell'$
must be both even or odd, and this means that $L$
must be even or else the spin--component
spectral function will be identically zero. Therefore, one has
$L=0$ and $\ell=\ell'$.
With these results one can easily see that
the dependence on the angular variables $\Omega=\{\theta,\phi\}$
disappears and the spectral function depends only on the magnitude
of the struck--nucleon momentum $p$. On the other hand, $m$ and
$m'$ must be equal. Finally,
using the relations with the 3--j and
6--j coefficients, one obtains the well--known result for the totally
unpolarized situation:$^{[2]}$
$$ S_{mm'}(p,E_m)=S(p,E_m)\delta_{mm'}=
{1\over 8\pi} {1\over [J_A]^2}
\sum_B\sum_{\ell j}
|C_{\ell j}(p)|^2\delta(E_m+\epsilon_B-\epsilon_A^0)\delta_{m m'}
\ \ .\eqno(A.11)$$

\vfill\eject
\smallskip
\centerline{\bf REFERENCES}
\medskip
\item{1.}
A. L. Dieperink and T. de Forest, {\it Ann. Rev. Nucl. Sci.\/} {\bf 25}
(1975) 1.
\medskip
\item{2.}
S. Frullani and J. Mougey, {\it Adv. in Nucl. Phys.\/} {\bf 14}
(1984).
\medskip
\item{3.}
J. A. Caballero, T. W. Donnelly and G. I. Poulis, {\it Preprint CTP--2046\/}.
{\it Nucl. Phys.} {\bf A} (in press); G. I. Poulis, {\it Ph. D. Thesis\/}
(MIT 1992, unpublished).
\medskip
\item{4.}
T. de Forest, {\it Nucl. Phys.} {\bf A392} (1983) 232.
\medskip
\item{5.}
C. M. Lederer and V. S. Shirley Eds., ``Table of Isotopes" (John Wiley $\&$
Sons, N.Y., 1978).
\medskip
\item{6.}
A. Bohr and B. R. Mottelson, {\it Mat. Fys. Medd. Dan. Vid. Selsk.\/}
{\bf 27} (1953) 16.; ``Nuclear Structure'' Vol. II (Benjamin, N.Y., 1975).
\medskip
\item{7.}
E. Moya de Guerra, {\it Phys. Rep.\/} {\bf 138} (1986) 293
and references therein.
\medskip
\item{8.}
E. Moya de Guerra, J. A. Caballero and P. Sarriguren, {\it Nucl. Phys.}
{\bf A477} (1988) 445.
\medskip
\item{9.}
D. Berdichevski {\it et al.}, {\it Phys. Rev.} {\bf C38} (1988) 338.
\medskip
\item{10.}
P. Sarriguren {\it et al.}, {\it Phys. Rev.} {\bf C40} (1989) 1414.
\medskip
\item{11.}
E. Moya de Guerra, {\it Phys. Rev.} {\bf C27} (1983) 2987; {\it An. de
F\'{\i}sica} {\bf A80} (1984) 32.
\medskip
\item{12.}
E. Garrido, E. Moya de Guerra, P. Sarriguren and J. M. Ud\'{\i}as,
{\it Nucl. Phys.} {\bf A550} (1992) 391.
\medskip
\item{13.}
J. A. Caballero and E. Moya de Guerra, {\it Nucl. Phys.}
{\bf A509} (1990) 117.
\medskip
\item{14.}
E. Moya de Guerra {\it et al.,} {\it Nucl. Phys.} {\bf A529} (1991) 68.
\medskip
\item{15.}
J.B.J.M. Lanen {\it et al.,} accepted in {\it Nucl. Phys.} {\bf A}.
\medskip
\item{16.}
A. S. Raskin and T. W. Donnelly, {\it Ann. Phys.\/} {\bf 191} (1989) 78.
\medskip
\item{17.}
T. W. Donnelly, ``New Frontiers in Electron Scattering: Spin Physics
and Coincidence Reactions," {\it Proc. of Nucl. Phys. Summer
School\/}, Oregon State University, Corvallis, OR (1988);
``Polarization in Electron--Scattering from Nucleons
and Nuclei,''  {\it Proc. of IV Jorge Andre Swieca Summer School in
Nuclear Physics\/}, Caxambu, Brazil (1989);
``Polarization Degrees of Freedom in Electron Scattering
from Nuclei," {\it Modern Topics in Electron Scattering\/}, Eds.
B. Frois and I. Sick (World Scientific, 1990); ``Special Topics in
Electromagnetic Nuclear
Physics and Related Aspects of Electroweak Interactions,''
{\it Proc. of VI Jorge Andre Swieca Summer School in
Nuclear Physics\/}, Campo do Jord\~ao, Brazil (1993).
\medskip
\item{18.}
H. W. L. Naus, S. J. Pollock, J. H. Koch and U. Oelfke,
{\it Nucl. Phys.\/} {\bf A509} (1990) 717 and references therein.
\medskip
\item{19.}
H. W. L. Naus and J. H. Koch,
{\it Phys. Rev.\/} {\bf C36} (1987) 2459; {\it Phys. Rev.\/}
{\bf C39} (1989) 1907.
\medskip
\item{20.}
P. C. Tiemeijer and J. A. Tjon, {\it Phys. Rev.\/} {\bf C42} (1990) 599.
\medskip
\item{21.}
X. Song, J. P. Chen, P. K. Kabir and J. S. McCarthy,
{\it J. Phys. G.: Nucl. Part. Phys.\/} {\bf 17} (1991) L75.
\medskip
\item{22.}
X. Song, J. P. Chen and J. S. McCarthy,
{\it Z. Phys.\/} {\bf A341} (1992) 275.
\medskip
\item{23.}
J. Mougey, {\it Ph. D. Thesis} (University of Paris--Sud 1976, unpublished).
\medskip
\item{24.}
J. Mougey {\it et al.}, {\it Nucl. Phys.\/} {\bf A262} (1976) 461.
\medskip
\item{25.}
A. R. Edmonds, ``Angular Momentum in Quantum Mechanics", (Princeton Univ.
Press, 1957).
\medskip
\item{26.}
E. Moya de Guerra, ``The Building Blocks of Nuclear Structure,'' Ed. A. Covello
(World Scientific, 1993).
\medskip
\item{27.}
S. G. Nilsson, {\it Mat. Fys. Medd. Dan. Vid. Selsk.} {\bf 29} (1955) 16.
\medskip
\item{28.}
P. Ring and P. Schuck, `` The Nuclear Many--Body Problem", (Springer--Verlag,
1980).
\medskip
\item{29.}
M. Vallieres and D. W. L. Sprung, {\it Can. J. Phys.\/} {\bf 56} (1978) 1190.
\medskip
\item{30.}
D. Vautherin and D. M. Brink, {\it Phys. Rev.} {\bf C5} (1972) 626;
D. Vautherin, {\it Phys. Rev.\/} {\bf C7} (1973) 296.
\medskip
\item{31.}
H. S. K\"{o}ler, {\it Nucl. Phys.\/} {\bf A258} (1976) 301.
\medskip
\item{32.}
S. Galster {\it et al.}, {\it Nucl. Phys.\/} {\bf B32} (1971) 221.
\medskip
\item{33.}
M. J. Musolf and T. W. Donnelly, {\it Nucl. Phys.\/} {\bf A546} (1992) 509;
E. Hadjimichael, G. I. Poulis and T. W. Donnelly, {\it Phys. Rev.\/}
{\bf C45} (1992) 2666.
\medskip
\item{34.}
D. B. Day, J. S. McCarthy, T. W. Donnelly and I. Sick,
{\it Ann. Rev. Nucl. and Part. Sci.} {\bf 40} (1990) 357.
\medskip
\item{35.}
M. Rotenberg, R. Bivins, N. Metropolis and J. K. Wooten,
``The 3--j and 6--j Symbols", (Technology Press, MIT, Cambridge, MA, 1959).
\medskip
\vfill\eject

\centerline{\bf FIGURE CAPTIONS}
\medskip
\item{Fig.~1:}Feynman diagram for the exclusive electromagnetic
${\svec A}({\svec e},e'N)B$
electron scattering process in the one--photon--exchange approximation.
\medskip
\item{Fig.~2:}Kinematics for two--arm coincidence reactions with polarized
targets.  ${\bf u}_z$ is along ${\bf q}$, ${\bf u}_y$ is normal to the
electron scattering plane and ${\bf u}_x = {\bf u}_y \times {\bf u}_z$ lies in
the scattering plane.  The target polarization vector
{\bf P}$^*$ is specified by the
angles $(\theta^*,\phi^*)$ in this coordinate system.  The unit vectors
${\bf u}_1$, ${\bf u}_2$, ${\bf u}_3$ are constructed by rotating the
previous ones (${\bf u}_x$, ${\bf u}_y$, ${\bf u}_z$)
by an angle $\phi_N$: ${\bf u}_3\equiv
{\bf u}_z$, ${\bf u}_1$ lies in the nucleonic plane and ${\bf u}_2$ is
perpendicular to it.
\medskip
\item{Fig.~3:}One--photon--exchange diagram for the reaction
${\svec A}({\svec e},e'N)B$
in the plane--wave impulse approximation (PWIA).
\medskip
\item{Fig.~4:}Monopole and quadrupole single--neutron densities in (a)
{\bf r}--space and (b) {\bf p}--space obtained from DDHF (solid), Nilsson
with N--admixtures (dashed) and Nilsson without N--admixtures (short--dashed).
The $b$ value for the Nilsson model calculations is $b=1.79$ in fm units.
\medskip
\item{Fig.~5:}The momentum distribution $n_0$ from Eq.~(25) for the
$^{21}{\overrightarrow{{\hbox{Ne}}}}$ to $^{20}$Ne (g.s.) transition with
the target polarized in the $z$--direction (along {\bf q}) and with
$p_y=0$ ({\it i.e.,\/} where the struck nucleon is found in the $xz$--plane:
$\phi=0 \leftrightarrow$ co--planar kinematics). DDHF single--particle
wave functions have been used.
\medskip
\item{Fig.~6:}As for Fig.~5, except now for the momentum distribution
$n_l$ from Eq.~(29a).
\medskip
\item{Fig.~7:}As for Fig.~5, except now for the momentum distribution
$n_s$ from Eq.~(29b).
\medskip
\item{Fig.~8:}As for Fig.~5, except now for the momentum distributions
$n_0 = n_n$ from Eqs.~(25) and (29c), respectively, in the situation
where the target polarization lies along the $y$--axis.
\medskip
\item{Fig.~9:} Unpolarized differential cross section
(Eq.~(37)) for the reaction
$^{21}{\hbox{Ne}}(\hbox{e},\hbox{e}'\hbox{n})^{20}$Ne(g.s).
Results correspond to different values of the scattering angle
$\theta_e$ and the azimuthal angle of the ejected neutron $\phi_N$,
obtained by using Nilsson (full) and DDHF (dashed) single--nucleon wave
functions.
\medskip
\item{Fig.~10:}Asymmetry ${\cal P}_{\Sigma}$ (Eq.~(39)) for the reaction
$^{21}{\overrightarrow{{\hbox{Ne}}}}({\hbox{e}},\hbox{e}'\hbox{n}
)^{20}$Ne (g.s.).
Results correspond to different values of $\phi_N$ and different orientations
of the target polarization defined by the angles within parentheses
$(\theta^*,\phi^*)$, in degrees.
\medskip
\item{Fig.~11:}Same as Fig.~10, but for the asymmetry
${\cal P}_{\Delta}$ (Eqs.~(40)) for the reaction
$^{21}{\overrightarrow{{\hbox{Ne}}}}(\vec{\hbox{e}},\hbox{e}'\hbox{n}
)^{20}$Ne (g.s.).  Results correspond to forward (a) and
backward (b) electron scattering angles. In each case six different situations
are shown defined by the target polarization direction $(\theta^*,\phi^*$) and
the value of $\phi_N$: i) $\phi_N=0^o$ and the target polarization
vector ${\bf P}^*$ oriented along the axes $z$ (full),
$x$ (short--dashed) and $y$ (dashed), and ii)
$\phi_N=90^o$ and ${\bf P}^*$ parallel to $z$ (long--dashed),
parallel to $x$ (dotted) and parallel to $y$ (dot--dashed).
\medskip
\item{Fig.~12:}Same as Fig.~9, except that now contributions from the first
excited state in the residual nucleus $^{20}$Ne are also included.
\medskip
\item{Fig.~13:}Same as Fig.~10, except now including the contributions from the
first excited state in the residual nucleus $^{20}$Ne. For the different
kinematical situations considered a comparison between the results obtained
with
Nilsson (full) and DDHF (dashed) single--nucleon wave functions is presented.
\medskip
\item{Fig.~14:}Same as Fig.~13, but now for the asymmetry ${\cal P}_{\Delta}$.
Results correspond to $\theta_e=30^o$ and different
orientations of the target polarization: (a) along $z$, (b) along $x$ and
(c) along $y$, and different values of $\phi_N$.
\medskip
\item{Fig.~15:}Same as Fig.~14, except now for backward--angle
electron scattering ($\theta_e = 150^\circ$).
\medskip
\item{Fig.~16:}Hadronic response functions $R^{K/K'}$, $K=L,T,TL,TT$,
$K'=T',TL'$ (see text). Contributions from the ground and
first excited state in $^{20}$Ne are included and the value of
$\phi_N$ is taken to be $\phi_N=0^o$.
A comparison between the results obtained with Nilsson and DDHF wave functions
for two orientations of the target polarization is presented:
i) ${\bf P}^*$ parallel to ${\bf q}$ with Nilsson (full) and DDHF (dashed)
wave functions, and ii)
${\bf P}^*$ parallel to the $x$--axis with Nilsson (short--dashed) and DDHF
(long--dashed) wave functions.
\medskip
\item{Fig.~17:}Asymmetry ${\cal P}_{\Sigma}$ versus the target polarization
angle $\theta^*$. Results correspond to $\phi^*=0^o$ and the residual
nucleus in its ground state ($g.s.$). Panel (a) corresponds to
$\phi_N=0^o$ and panel (b) to $\phi_N=90^o$. In the two cases
we show the results obtained for five different values of
the struck--neutron momentum:
$p=0.2$ fm$^{-1}$ (full), 0.6 fm$^{-1}$ (short--dashed), 1.0 fm$^{-1}$
(dashed), 1.4 fm$^{-1}$ (dot--dashed) and 1.8 fm$^{-1}$ (dotted).
\medskip
\item{Fig.~18:}Same as Fig.~17, except that now we also include the
contributions from the first excited state in the residual nucleus $^{20}$Ne.
\medskip
\item{Fig.~19:}Same as Fig.~17, but including the first two excited states of
the daughter nucleus $^{20}$Ne.
\medskip
\item{Fig.~20:}Some as Fig.~17, except now for the asymmetry
${\cal P}_{\Delta}$ and the scattering angle $\theta_e=30^o$.
\medskip
\item{Fig.~21:}Same as Fig.~20, except now including the contributions from the
states $0^+$ and $2^+$ in $^{20}$Ne.
\medskip
\item{Fig.~22:}Same as Fig.~21, now also including the contributions from the
second excited state $4^+$ in $^{20}$Ne.
\medskip
\item{Fig.~23:}Differential cross section (Eq.~(1)) versus the target
polarization angle $\theta^*$. Results correspond to
$p=0.75$ fm$^{-1}$ and $\theta_e=30^o$.
DDHF single--nucleon wave functions have been used and
contributions from the ground and
first excited state in the residual nucleus $^{20}$Ne are included.
Panel (a) corresponds to $\phi_N=0^o$, panel (b) to
$\phi_N=66^o$ and panel (c) to $\phi_N=90^o$. In each case the results obtained
for different values of $\phi^*$ are represented. In panels (a) and (c)
$\phi^*=0^o$ (full), $\phi^*=30^o$ (short--dashed), $\phi^*=60^o$ (dashed) and
$\phi^*=90^o$ (dot--dashed), while in panel (b) $\phi^*=6^o$ (full),
$\phi^*=26^o$ (short--dashed),
$\phi^*=46^o$ (dashed), $\phi^*=66^o$ (dot--dashed) and
$\phi^*=86^o$ (dash--double--dotted). Also for comparison, in the three graphs
the results for the totally unpolarized cross section (Eq.~(37)) are shown
(dotted).
\item{Fig.~24:}Unpolarized cross section (a)
and asymmetries ${\cal P}_{\Sigma}$/${\cal P}_{\Delta}$ (b)
versus the nuclear deformation
parameter $\delta$ divided by the equilibrium results ($\delta_{eq}=0.31$).
Results correspond to forward--angle electron scattering
$\theta_e=30^o$ and co--planar kinematics $\phi_N=0^o$.
Contributions from the ground and first excited state
in the residual nucleus $^{20}$Ne are included. Results are shown for
various values of
the struck--neutron momentum: $p=0.5$ fm$^{-1}$ (full),
0.75 fm$^{-1}$ (short--dashed) and 1.5 fm$^{-1}$ (dashed) in case (a),
and $p=0.75$ fm$^{-1}$ for ${\cal P}_{\Sigma}$ (solid), ${\cal P}_{\Delta}$
(dashed), and $p=1.5$ fm$^{-1}$ for ${\cal P}_{\Sigma}$
(short--dash), ${\cal P}_{\Delta}$ (long--dash).
\medskip
\item{Fig.~25:}Ratio between the results obtained for $q=1$ GeV and
$q=500$ MeV at the corresponding quasielastic peaks and $\theta_e=30^o$.
DDHF single--nucleon wave functions have been used and contributions from
the ground and first excited state in $^{20}$Ne have been included.
Panel (a) corresponds to the totally
unpolarized cross section (Eq.~(37)) with results for $\phi_N=0^o$ (full) and
$\phi_N=90^o$ (dashed). Panel (b) corresponds to the asymmetry
${\cal P}_{\Sigma}$, where the three curves shown correspond to $\phi_N=0^o$
and
target polarization along $x$ (full), $y$ (dashed) and $z$ (dotted).
Finally, panel (c) shows the ratio for the asymmetry ${\cal P}_{\Delta}$
for, i) $\phi_N=0^o$ and ${\bf P}^*$ parallel to
$x$ (full), parallel to $z$ (dashed), and ii) $\phi_N=90^o$ with
${\bf P}^*$ parallel to $x$ (dotted), parallel to $y$ (short--dashed)
and parallel to $z$ (dot--dashed).

\goodbreak\bigskip
\centerline{\bf TABLE CAPTIONS}
\medskip
\item{Table 1:}Choices of target polarization and outgoing nucleon directions
where the electron--polarized response functions
$R^{T'}$ and/or $R^{TL'}$ vanish.
\medskip
\item{Table 2:}Quadrupole moments, r.m.s. radii and $\beta$--values
in {\bf r} and {\bf p}--spaces for neutrons ($\nu$) and protons ($\pi$)
obtained with Nilsson and density dependent DDHF calculations (see text).
\medskip
\item{Table 3:}Weights $n_{\ell j}$ of the projected angular momentum
components of the odd--neutron wave function in $^{21}$Ne. Nilsson
and DDHF results are shown (values lower than $10^{-3}$ have
been omitted).
\medskip
\item{Table 4:}Values of the angles $\theta$ and $\theta_N$ that correspond to
the five $p$--values used in the calculations (see Eqs.~(51) and (52)).
The residual nucleus $^{20}$Ne is considered in its ground state.
\medskip
\vfill
\eject
\end